\documentclass{JHEP3}
\usepackage{amssymb}
\usepackage{amsfonts}
\usepackage{amsbsy}
\usepackage{amsmath}
\usepackage{amsthm}
\usepackage{graphicx}
\usepackage{epstopdf}
\usepackage[vcentermath]{youngtab}
\usepackage{multirow}
\usepackage{latexsym}
\usepackage{array}
\usepackage{epsfig}

\def\half{\frac{1}{2}}
\def\Z{\mathbb{Z}}
\def\F{\mathbb{F}}

\def\cL{\mathcal{L}}
\def\R{\mathbb{R}}

\def\P{\mathbb{P}}
\def\Tr{{\mathrm{Tr}}}
\def\tr{{\mathrm{tr}}}

\def\ho{h^{1, 1}}
\def\htb{H_2 (B,\Z)}
\def\tf{\tilde{\F}}

\def\cs{{C_{\rm cs}}}
\def\csa{{C_{\rm sa}}}
\def\cp{{C_{\rm 21}}}
\def\rp{{r_{\rm 21}}}

\newcommand{\cO}{\mathcal{O}}

\newcommand{\cV}{\mathcal{V}}

\newcommand{\cC}{\mathcal{C}}

\newcommand{\cN}{\mathcal{N}}

\newcommand{\cA}{\mathcal{A}}

\newcommand{\cB}{\mathcal{B}}
\newcommand{\cF}{\mathcal{F}}

\newcommand{\cR}{\mathcal{R}}

\newcommand{\beq}{\begin{equation}}
\newcommand{\eeq}{\end{equation}}
\newcommand{\be}{\begin{equation}}
\newcommand{\ee}{\end{equation}}
\newcommand{\bea}{\begin{eqnarray}}
\newcommand{\eea}{\end{eqnarray}}   
\newcommand{\ben}{\begin{eqnarray*}}
\newcommand{\een}{\end{eqnarray*}}                  
\newcommand{\ba}{\begin{aligned}}
\newcommand{\ea}{\end{aligned}}
\newcommand{\bt}{\begin{tabular}}
\newcommand{\et}{\end{tabular}}
\newcommand{\bc}{\begin{center}}
\newcommand{\ec}{\end{center}}

\newcommand{\eq}[1]{(\ref{#1})}

\newcommand{\drop}[1]{}

\title{Structure in 6D and 4D ${\cal N} = 1$ supergravity theories
from F-theory}

\author{Thomas W.~Grimm$^{1}$ and Washington Taylor$^2$\\
$^1$Max Planck Institute for Physics\\
          F\"ohringer Ring 6  \\
          Munich, 80805, Germany \\

$^2$Center for Theoretical Physics\\
Department of Physics\\
Massachusetts Institute of Technology\\
Cambridge, MA 02139, USA\\
\\
\\
{\tt grimm} {\rm at} {\tt mppmu.mpg.de},
{\tt wati} {\rm at} {\tt mit.edu}
}

\preprint{MPP-2012-72\\
MIT-CTP-4248}

\abstract{We explore some aspects of 4D supergravity theories and
  F-theory vacua that are parallel to structures in the space of 6D
  theories.  The spectrum and topological terms in 4D supergravity
  theories correspond to topological data of F-theory geometry, just
  as in six dimensions.  In particular, topological axion-curvature
  squared couplings appear in 4D theories; these couplings are
  characterized by vectors in the dual to the lattice of axion shift symmetries
  associated with string charges.  
  These terms are analogous to the
  Green-Schwarz terms of 6D supergravity theories, though in 4D the
  terms are not generally linked with anomalies.  We
  outline the correspondence between F-theory topology and data of the
  corresponding 4D supergravity theories.  The correspondence of
  geometry with structure in the low-energy action illuminates
  topological aspects of heterotic-F-theory duality in 4D as well as
  in 6D.  The existence of an F-theory realization also places
  geometrical constraints on the 4D supergravity theory in the large-volume limit.
}

\begin{document}

\section{Introduction}
\label{sec:intro}

F-theory \cite{Vafa-F-theory, Morrison-Vafa-I, Morrison-Vafa-II}
provides a very general approach to constructing string vacua in
even-dimensional space-times.  In particular, F-theory gives a
nonperturbative description of a wide range of string
compactifications.  F-theory describes structures such as gauge
groups, matter fields, and Yukawa couplings in a simple geometric
framework that is amenable to the use of powerful mathematical tools
from algebraic geometry.  
F-theory as it is currently understood is incomplete as a physical theory.
In its elemental geometric formulation there is no action principle 
or complete characterization of the fundamental degrees of freedom.
The clearest definition of F-theory is as a limit of M-theory.
M-theory itself, however, is also not a completely well defined
theory, and some of the mathematical simplicity of F-theory is less
apparent in the M-theory framework.
Nonetheless, even with its current limitations, F-theory has proven to
be a powerful tool for exploring both the large-scale structure of the
landscape of string vacua and detailed aspects of semi-realistic
phenomenology.

In eight and six dimensions, the set of F-theory compactifications
includes vacua with spectra matching those of most or all
supergravity theories that can be realized using other known string
theory constructions (for a review of 8D and 6D supergravity/F-theory
models and many further references, see \cite{WT-TASI}).  Given a
six-dimensional supergravity theory, the spectrum and action of the
theory provide data that can be used to identify the geometry of a
corresponding F-theory construction, when one exists 
\cite{Sadov, KMT,KMT-II,BonettiGrimm}.  The F-theory geometry in turn imposes certain constraints
on the spectrum and action of the low-energy theory
\cite{Grassi-Morrison, KMT-II, Seiberg-Taylor, Grassi-Morrison-2}.
Some, but not all, of these constraints are understood from
macroscopic/low-energy consistency conditions such as anomaly
cancellation.  The set of 6D F-theory vacua forms a complicated moduli
space with many components associated with different F-theory ``base''
geometries connected through extremal tensionless string transitions
\cite{Seiberg-Witten, Morrison-Vafa-II}.  Recent work has begun to
systematically classify the set of 6D F-theory compactifications,
using connections between the F-theory geometry and corresponding
structure in the low-energy supergravity theory \cite{bound, KMT-II,
  0, Morrison-Taylor-matter, Braun, Morrison-Taylor, mt-toric}.

In four dimensions F-theory gives rise to an even broader and richer
class of vacua than in higher dimensions. Recent efforts have 
focused on compactifications relevant for semi-realistic GUT phenomenology \cite{Heckman-review,Weigand-review}
including constructions of compact fourfolds for global models \cite{Marsano-F-theory,Blumenhagen-F-theory,Cvetic-gh,Knapp:2011wk}.  
In four dimensions,
however, with only one supersymmetry, the space of string solutions is
complicated by various perturbative and nonperturbative effects such
as fluxes that remove massless moduli and produce a ``landscape''
containing isolated distinct vacua connected through regions of
off-shell string physics (for reviews of flux compactifications and
related developments see {\it e.g.} \cite{Douglas-Kachru,Blumenhagen:2006ci,Denef:2008wq}).  In this context, the limitations of F-theory in
its current form become more apparent, and using this nonperturbative
approach to study the global space of solutions becomes more
challenging.  Recent work has focused on incorporating more directly into F-theory
degrees of freedom such as fluxes on the
world-volume of 7-branes 
\cite{Grimm-hkk,Braun:2011zm,Marsano:2011hv,Krause:2011xj,Grimm:2011fx,Krause:2012yh,Intriligator-jmmp}, 
and the
related transverse scalar fields on multiple branes that can carry
noncommuting structures such as ``T-branes'' \cite{T-branes}.  While
these features are present in the M-theory description
of F-theory, the 4D physics described by F-theory is only reached in a singular 
limit that is as yet not fully understood. In four dimensions,
there also appear to be many types of string solutions that are not
easily described in the F-theory framework, such as $G_2$
compactifications of M-theory \cite{Duff:1986hr,Papadopoulos:1995da,Acharya:2001gy}, heterotic compactifications
on Calabi-Yau manifolds that (unlike K3) have no elliptic fibration in
their moduli space, and other more exotic possibilities that may
include a vast range of asymmetric orbifolds
\cite{ao}  and/or  non-geometric flux vacua \cite{hmw,
  Dabholkar-Hull, Kachru-stt,Hull,
  stw, Wecht-review} (that may also have  asymmetric orbifold descriptions \cite{Condeescu:2012sp}).  Despite these limitations, it can be
argued that at this stage F-theory provides the broadest perspective
on the range of possible phenomena that may emerge from string theory
in 4D supergravity theories.  In this work we address some global
questions regarding the structure of F-theory vacua within the
existing framework. For many practical questions  we use the 
definition of F-theory as a limit of M-theory as recently studied in the 4D context
in \cite{Denef:2008wq, Grimm-F-theory}.

In six dimensions, the key to reconstructing the geometry of an
F-theory compactification from the data of the supergravity theory
lies in the Green-Schwarz terms of the form $B R^2$ and
$BF^2$, and in the related lattice of dyonic string charges.  While
the original understanding of the Green-Schwarz terms arose through
the anomaly cancellation mechanism, it seems that there may be deeper
reasons underlying the existence and structure of these terms.  
In four dimensions similar topological
couplings arise between axions
$\rho$ and gauge and gravitational
curvature-squared terms, of the form
\begin{equation}
(a \cdot \rho) R \wedge R + \sum_{A} (b_A \cdot \rho) F^A \wedge F^A \,.
\label{eq:4D-ab}
\end{equation}
While in some cases these terms are connected with a generalized
Green-Schwarz mechanism for cancellation of gauge and mixed
abelian-gravitational anomalies \cite{Green:1984sg,Blumenhagen:2005mu}, these terms are
not uniquely determined by this condition; for example, the $\rho R
\wedge R$ terms appear even in theories without massless gauge fields. 
A detailed discussion of the  generalized
Green-Schwarz mechanism in related weakly coupled Type IIB scenarios
can be found in \cite{Plauschinn}.
Terms of the form (\ref{eq:4D-ab}), and the associated integral
lattice of axionic string/instanton charges (containing $a$ and $b_A$
in \eq{eq:4D-ab}), relate 4D supergravity theories to F-theory
geometry in a parallel fashion to the six-dimensional story.  In
particular, $a$ contains geometric information about the F-theory
compactification manifold (the canonical class of the threefold base),
while $b_A$ captures information about the geometric structure giving
rise to the simple factors in the gauge group (the locations of the
7-branes supporting the gauge group factors).  
As in six dimensions, this information, along with 
other structure in the 4D supergravity theory,
can be used in a ``bottom-up''
fashion to identify  the F-theory geometry
needed for a UV completion of the theory.
We describe in this paper how the terms
(\ref{eq:4D-ab}) arise from F-theory, and match with dual heterotic
constructions.  This connection gives a simple perspective on the
topological structure of heterotic-F-theory duality that is valid for
$SO(32)$ as well as $E_8 \times E_8$ heterotic vacua with F-theory
duals.
More generally, these couplings and the structure of the related string
charge lattice may provide a useful tool for addressing global
questions about the space of string vacua and related duality
symmetries in 4D just as they have done in 6D.

Some previous progress towards relating the degrees of freedom and
action of 4D ${\cal N} = 1$ supergravity theories to the data used in
an F-theory construction via M-theory was presented in \cite{Grimm-F-theory,Grimm:2011tb,Grimm:2011sk,Grimm:2012rg}.  In
four dimensions, the structure of F-theory compactifications is
complicated by the necessary presence of fluxes that produce a
superpotential or D-terms that lift some moduli of the theory.  In this paper,
we assume that the theory is in a regime where these moduli are light,
corresponding to a large-volume F-theory compactification.  In this
regime, F-theory geometry places certain constraints on the spectrum
and action of the associated 4D supergravity theory.  An important
direction for further extension of the work in this paper is to
develop an understanding the role of the structure and constraints
presented here away from the large-volume F-theory limit.

Six-dimensional supergravity theories and F-theory vacua are described
in Section \ref{sec:6D}.  The spectrum and relevant terms in the
action of 4D theories are described in Section \ref{sec:4D}.  This
section also contains an analysis of axion--curvature squared
couplings in heterotic theories, and uses these terms to determine
topological aspects of the general heterotic/F-theory duality correspondence for
4D theories.  Section \ref{sec:constraints} contains a brief
description of some structures  and constraints
on 4D theories
associated with large-volume F-theory compactifications that are close
analogues of similar structures and constraints in six dimensions.
Section \ref{sec:conclusions} contains concluding
remarks.

\section{Six-dimensional supergravity theories and F-theory vacua}
\label{sec:6D}

In this section we summarize some key features of 6D supergravity
theories and F-theory vacua.  We outline the correspondence between
data in the supergravity theory and geometric structures in F-theory.
Most of this material is known and is
described in earlier papers, but here we consolidate together a
variety of results on 6D theories into a coherent picture for
comparison with the 4D story.  We also add a few new observations on
some aspects of 6D theories that help to clarify both the 6D and 4D
stories.
The material in this section is also used in Section \ref{sec:constraints} to
characterize constraints on consistent 6D F-theory vacua in terms of data
in the supergravity theory.

\subsection{F-theory vacua and 6D spectra}
\label{sec:6D-basics}

A review of 6D supergravity theories and associated F-theory
constructions appears in \cite{WT-TASI}; another review of 6D string
vacuum constructions from a variety of approaches including F-theory
is given in \cite{Ibanez-Uranga}.  We summarize the basics here,
beginning with supergravity and then describing F-theory models.

We begin with some generalities on 6D 
supergravity theories with $\cN=(1,0)$ supersymmetry,
corresponding to eight supercharges.
The spectrum of such a theory contains:
\begin{itemize} 
\item One gravity multiplet,\\[-.8cm]
\item $T$ tensor multiplets,\\[-.8cm]
\item $V$ vector multiplets in a general (nonabelian $\times$
abelian) gauge group,\\[-.8cm]
\item  $H$ hypermultiplets spanning a 
quaternionic K\"ahler manifold.
\end{itemize} 
Each hypermultiplet contains four real scalars as bosonic components.
Note that the bosonic components of the gravity multiplet contain, in
addition to the metric, a two-form field with self-dual field
strength.  The tensor multiplets each contain an anti-self-dual
two-form field as well as a real scalar.  
The field content, couplings, and equations of motion of 6D
supergravity theories were studied in \cite{Nishino-Sezgin, gsw,
  Romans, Sagnotti, Ricci-Sagnotti}.  
Dyonic strings in the theory
carry charges under the self-dual and anti-self-dual two-form fields
in the gravity and tensor multiplets.  These strings should appear in any
quantized theory of 6D supergravity as quantum excitations charged
under the two-form fields, independent of whether the UV completion of
the gravity theory involves a conventional formulation of string theory.
The charges of the quantized dyonic strings lie
in a lattice $\Gamma$ that must be unimodular \cite{Seiberg-Taylor}.

In six dimensions, anomaly cancellation \cite{gsw, Sagnotti, Erler}
provides a powerful set of
constraints on the set of possible theories, as well as a useful tool
for analyzing supergravity theories.  For example, the numbers of
multiplets $H,V,T$ introduced above are not independent, but rather
linked through the gravitational anomaly relation
\begin{equation} \label{eq:HVtext}
     H-V =273-29T\, .
\end{equation} 
We briefly review the complete set of 6D anomaly cancellation
conditions in Appendix \ref{6Danomalies}.

We now turn to F-theory constructions of 6D ${\cal N} = (1, 0)$
supergravity theories.  Such models arise from
compactification of F-theory on an elliptically fibered Calabi-Yau
threefold $X$ over a complex surface base $B$.  A detailed description
of 6D compactifications of F-theory is given in \cite{Morrison-Vafa-I,
  Morrison-Vafa-II}; we briefly review the structure of these vacua,
emphasizing the correspondence between the topology of $B, X$ and the
field content of the low-energy theory.  

In the type IIB picture, an F-theory construction is given
by a set of 7-branes wrapped on the space $B$.  
The nonabelian part of the gauge group arises from
coincident 7-branes on $B$, which give singularities in the
elliptic fibration $X$ associated with codimension one loci
(divisors) on the base. The nonabelian gauge group factor on such a
divisor can be determined from the Kodaira/Tate classification of the
local codimension one singularity \cite{Morrison-Vafa-I,
  Morrison-Vafa-II, Bershadsky-all, Morrison-sn,
Grassi-Morrison-2}.  
The elliptically
fibered threefold $X$  is given in the Weierstrass description by
\begin{equation}
y^2 = x^3 +
fx + g,
\label{eq:Weierstrass}
\end{equation} 
where $f, g$ are sections of $-4K, -6K$, with $K$ being the 
canonical bundle of the base $B$. 
In the type IIB picture, the 7-branes are wrapped on 
the two-cycles in the base where the elliptic fibration degenerates.
This degeneration locus is given by the vanishing of the discriminant $\Delta = 4f^3 + 27g^2$.
The Kodaira condition that the
total space of the elliptic fibration be Calabi-Yau states that
\begin{equation}
- 12 \, [K] = [\Delta] = \sum_{A} \nu_A [S_A] + [Y] \,
\label{eq:Kodaira}
\end{equation}
where $[K]=-c_1(B)$ is the canonical class of the base, $[\Delta]$ is
the total class of the singularity locus, $[S_A]$ are the classes of
the irreducible effective divisors carrying simple gauge group factors
$G_A$, and $Y$ is the residual discriminant locus, which does not give
rise to nonabelian gauge symmetries.  The divisors $S_A$ carrying
nonabelian gauge group factors are associated with singularities in
the fibration characterized by integer multiplicities $\nu_A$
depending on the group $G_A$.  ({\it e.g.}, $\nu = N$ for $SU(N)$,
$\nu = 10$ for $E_8$, {\it etc.})
We use Poincar\'{e} duality to move
freely between divisor classes in $H_2 (B)$ and elements of $H^{1, 1} (B)$.  Two
viewpoints on such compactifications will be useful. Either we can
consider the base $B$ supplemented by additional data for the 7-branes
or we can study the complete singular threefold $X$.

We can now describe how the 6D spectrum is related to the F-theory
geometry.
The number of tensor
multiplets is
related to the topology of the F-theory base through
\begin{equation}
T = h^{1, 1} (B) -1 \,.
\label{eq:t-h11}
\end{equation}
There is a unimodular lattice $\Gamma = H_2 (B,\Z)$ of dyonic string
charges of signature $(1, T)$ associated with type IIB D3-branes
wrapped on the 2-cycles of $B$.  These strings are charged under the
self-dual and anti-self-dual two-form fields in the theory.

The number of vector fields depends not only on the topology of the
F-theory base but also on the singularity structure of the fibration
encoded in $\Delta$.
These singularities will generically render the total space $X$ of the
elliptic fibration  singular. 
One can, however, canonically 
blow up the singularities at each codimension, producing a smooth 
Calabi-Yau space $\hat X$.
The rank of the gauge group of the 6D theory is then given  in terms
of the topologies of $\hat{X}$ and $B$ by
\begin{equation}
r = h^{1, 1} (\hat X) -h^{1, 1} (B) -1 \,.
\label{eq:6D-rank}
\end{equation}
This rank can include a number of abelian vector fields in addition to
the nonabelian gauge fields.  Abelian vector fields are associated
with extra sections of the fibration that increase the rank of the
Mordell-Weil group \cite{Morrison-Vafa-II}; the treatment of such
abelian factors is rather subtle.

Finally, the number of uncharged scalar fields is
\begin{equation}
H_{\rm neutral} = h^{2, 1} (\hat X) + 1 \,.
\label{eq:6D-neutral}
\end{equation}
These fields come from the complex structure moduli on $X$, with one
modulus for the overall K\"ahler class of the base $B$; equivalently,
these fields correspond to physical moduli in the Weierstrass
description \eqref{eq:Weierstrass} of the F-theory model\footnote{A
  detailed counting of physical vs.~non-physical degrees of freedom in
  the Weierstrass coefficients of 6D F-theory models appears in
  \cite{mt-toric}.}.  (More precisely, 
the fields $H_{\rm neutral}$ are quaternionic,
with four real degrees of freedom in each field; half of the degrees
of freedom in each neutral field other than the overall K\"ahler modulus come
from complex structure/Weierstrass moduli, the other half come from
degrees of freedom on the 7-branes and the bulk form fields.)  In general, charged matter
fields arise from codimension 2 singularities in the elliptic
fibration \cite{Bershadsky-all, Katz-Vafa, Sadov,
  Morrison-Taylor-matter, Grassi-Morrison-2}, with some matter fields
such as adjoint representations arising nonlocally on divisors $S_A$
of higher genus \cite{enhanced, Witten-phase-mf}. Throughout this work
we will primarily focus on 6D theories without matter fields for
simplicity; in many theories there is a phase in which all matter
fields are Higgsed and there is no massless charged matter
\cite{Morrison-Taylor}.

Note that the identifications \eqref{eq:6D-rank}, \eqref{eq:t-h11} and
\eqref{eq:6D-neutral} allow us to give a simple expression for the
Euler character of the resolved threefold for theories without charged matter
\begin{equation} \label{generalchi_6D}
   \chi(\hat X) = 2 (h^{1,1}(X) - h^{2,1}(X)) = 2 (r +  T -  H_{\rm neutral}+ 3)
\end{equation}
This equation provides the simplest link between the topology of $\hat
X$ and the 6D spectrum.  As we discuss later,
$\chi(\hat X)$ can also be related to the constant coefficients determining the 
6D Green-Schwarz terms.  This connection between the topology of the F-theory
compactification space and the structure of the supergravity spectrum
and action provides a constraint on 6D supergravity theories that, as
we discuss further in Section \ref{sec:constraints}, matches with
6D anomaly cancellation conditions.

\subsection{F-theory geometry and terms in the 6D supergravity action}

In the previous section we described the correspondence between
topology of the F-theory compactification space and the spectrum of
the 6D theory.  We now consider the connection between further
geometric structures of the F-theory compactification and terms in the
supergravity action.  In particular, some terms in the supergravity
action carry discrete geometric information about the F-theory
picture.  Through understanding this correspondence we can construct
a map from data in the supergravity spectrum and action to data of the
F-theory geometry, allowing us to identify which specific F-theory
vacuum should correspond to any given supergravity theory.  As we
describe in Section \ref{sec:6D-heterotic}, this gives a simple way of
describing dualities such as heterotic/F-theory duality at the level
of topology.
Using the correspondence in the
opposite direction, we can interpret constraints associated with
F-theory geometry as necessary conditions for a supergravity theory to
admit an F-theory realization, as we discuss in Section \ref{sec:constraints}.

\subsubsection{Couplings in the 6D supergravity action}

As discussed above, from the spectrum of a given 6D theory one can
already infer some core topological data of the base $B$ and total
space $X$ of the F-theory elliptic fibration using \eqref{eq:t-h11},
\eqref{eq:6D-rank}, and \eqref{eq:6D-neutral}.  More precise data from
the supergravity theory is required to construct information about
specific divisor classes on the base $B$, such as the canonical class
$K$ of $B$, and the divisor classes $S_A$ carrying the nonabelian
gauge group factors.  This information is carried in the structure of
the Green-Schwarz terms of the 6D supergravity action, which take the
schematic form
\begin{equation}
(K \cdot B) \wedge R \wedge R, \;\;\;\;\;
(S_A \cdot B) \wedge F^A \wedge F^A
% \label{eq:}
\end{equation}
(where $B$ is the vector of two-forms).  We now describe these terms
in further detail.  The analogous structure in four dimensions is one
of the main focal points  of this paper.

The study of the action for 6D effective supergravity theories is
complicated by the fact that one has to deal with self-dual and
anti-self-dual two-forms $B_2^\alpha$.  This problem can be
overcome, however, by working with a pseudoaction, where the duality
constraints are imposed by hand after determining the equations of
motion \cite{Nishino-Sezgin,Ricci-Sagnotti}, as done for F-theory in \cite{BonettiGrimm}.  We focus here on the terms
that are needed to identify the internal geometry.  We take the Einstein-Hilbert term to have the canonical normalization
$S_{\rm EH} = - \int \tfrac12 R_s * \mathbf{1}$.  The terms in the
action in which we are particularly interested are the quadratic terms
in the  space-time, nonabelian, and two-form curvatures \footnote{In this action we have included a term $\text{tr}\, R \wedge * R $ in analogy to $\text{tr}\, F^A \wedge * F^A$.
The precise form of this term can be altered by a field redefinition
involving the metric. In F-theory these higher derivative terms 
can be determined via 5D M-theory compactifications generalizing~\cite{BonettiGrimm}. In five dimensions the 
supersymmetric completion of the curvature squared terms is known \cite{Hanaki:2006pj}.}
\bea \label{6Daction} S^{(6)} &=&
- \int \frac{1}{2} \left(j^\alpha \Omega_{\alpha \beta} a^\beta \right)
\ \text{tr}\, R \wedge * R + 
\frac{2}{\lambda_A}\left(j^\alpha \Omega_{\alpha \beta}
b^\beta_A \right) \   \text{tr}\, F^A \wedge * F^A  \\ 
&& \phantom{-\int} +\frac{1}{4} \left(B^\alpha \Omega_{\alpha \beta} a^\beta \right)\wedge \text{tr}\, R \wedge  R + \frac{1}{\lambda_A}
\left(B^\alpha \Omega_{\alpha \beta}
b^\beta_A \right) \wedge  \text{tr}\, F^A \wedge F^A \nonumber \\
&& \phantom{-\int} + \frac14 G_{\alpha \beta} H_3^\alpha \wedge * H_3^\beta + \frac12 G_{\alpha \beta} dj^\alpha \wedge * dj^\beta \ . \nonumber \eea 
Here $R = \frac12 R_{\mu \nu} dx^\mu \wedge dx^\nu$ is the
$SO(1,5)$-valued curvature two-form. We have introduced the field
strengths $H_3^\alpha$ of the $B_2^\alpha$ that are given by  
\begin{equation} \label{H3-correct}
   H_3^\alpha = d B_2^\alpha + \frac12 a^\alpha w_{\rm CS}(R)+2 \frac{b^\alpha_A}{\lambda_A} w_{\rm CS}^A(F)\ , 
\end{equation}
where the Chern-Simons forms are given by 
\bea \label{Chern-Simons-Forms}
  w_{\rm CS}(R) &=& \text{tr}\Big(\hat \omega \wedge d \hat \omega + \frac32  \hat \omega \wedge \hat \omega \wedge \hat \omega \Big) \ ,\qquad  \\ 
  w^B_{\rm CS}(F) &=& \text{tr} \Big(A^B \wedge d A^B + \frac32  A^B \wedge A^B \wedge A^B \Big) \ ,
\eea
with $\hat \omega$ being the spin-connection one-form.
The field $j^\alpha$ and the coefficients $a^\alpha,
\ b_A^\alpha$ transform as vectors in the space $\R^{1,T}$, which
carries a
symmetric inner product $\Omega_{\alpha\beta}$ of signature $(1, T)$.  
Note that the self- and anti-self duality conditions for $H_3^\alpha$
must be imposed by hand on the level
of the equations of motion by demanding 
\begin{equation} \label{self-dual_tensors}
   \Omega_{\alpha \beta} * H_3^\beta   = G_{\alpha \beta} H_3^\beta\ .
\end{equation}  
The field $j^\alpha$ contains the scalars in the $T$ tensor
multiplets. The additional degree of freedom in $j^\alpha$ is fixed by
the condition $j^\alpha \Omega_{\alpha \beta} j^\beta=1$. 
By convention, in (\ref{6Daction})
``$\tr$'' of $(F^B)^2$
denotes the trace in the fundamental representation, and $\lambda_B$
are normalization constants depending on the type of each simple group
factor.  
These constants are related to the dual Coxeter numbers $c_{G_B}$
of the gauge
group factors $G_{B}$ and trace normalization factors $A^{(B)}_{\rm adjoint}$ through $\lambda_B
= 2  c_{G_B}/A^{(B)}_{\rm adjoint}$, with coefficients discussed in
Appendix \ref{6Danomalies}.

The first two terms on the second line of (\ref{6Daction})
can be written as
\begin{equation}
 S^{(6)}_{\rm GS} =- \frac{1}{2} \int  \Omega_{\alpha
  \beta}\, B_2^\alpha \wedge X_4^\beta
\label{GS-term}
\end{equation}
where
\begin{equation} \label{GS-term-x}
   X_4^\alpha = \half a^\alpha \tr R \wedge R +  \sum_A b_A^\alpha \ \left(\frac{2}{\lambda_A} \tr F^A \wedge F^A \right)\ .
\end{equation}

For a 6D supergravity theory arising from F-theory, the $T+1$ two-form
fields $B_2^\alpha$ arise by expanding the R-R four-form $C_4$ of Type
IIB into a basis $\omega_\alpha$ of $h^{1,1}(B)$ two-forms spanning
$H^2(B)$ as $C_4= B^\alpha_2 \wedge \omega_\alpha$. The 6D tensors 
satisfy the duality condition \eqref{self-dual_tensors} due to the 10D self-duality of the 
field strength  $F_5$ of $C_4$.
Due to
the varying dilaton, however, this decomposition cannot in general be described
in a weakly coupled supergravity limit.  The 6D action can be derived
in a more precise fashion via a 5D M-theory compactification~\cite{BonettiGrimm}.  We now discuss the various terms appearing in
\eqref{6Daction} and comment on the topological information of the F-theory
compactification space that we can extract from these terms.

\subsubsection{Topological couplings in the Green-Schwarz term}
\label{sec:6D-gs}

We now discuss the terms from \eqref{6Daction} that appear
in (\ref{GS-term}).  
Equation (\ref{GS-term}) describes the 6D Green-Schwarz
terms that are needed for anomaly
cancellation \cite{gsw, Sagnotti, Erler}. 
In 6D there are  gauge, gravitational, and mixed gauge-gravitational
anomalies.  These anomalies are captured by an 8-form anomaly
polynomial $I_8(R,F)$ that is a
function of the curvature tensor $R$ and the gauge field strengths
$F^A$ of all gauge groups. 
If this polynomial factorizes as $I_{8} =
\frac12 \Omega_{\alpha \beta} X_4^\alpha X_4^\beta$, then the anomaly
can be cancelled using the Green-Schwarz counterterm, as described in
detail in Appendix B. 

For 6D supergravity theories arising from an F-theory compactification,
the $SO(1, T)$ vectors $a^\beta, b_A^\beta$ appearing in the 6D
Green-Schwarz terms carry topological information about the F-theory
geometry.  These vectors correspond to the canonical class and divisor
classes carrying the nonabelian gauge group factors in the F-theory
picture.  Specifically,
\begin{equation}
   a^\alpha = K^\alpha\ ,\qquad b^\alpha_A =  C^\alpha_A\ ,
\label{eq:Green-Schwarz-match}
\end{equation}
where the coefficients $K^\alpha$ and $C_A^\beta$ arise in the two-form expansions
\begin{equation} \label{Kxi-expansion}
    [K] = K^\alpha \omega_\alpha \ , \qquad [S_A] = C^\alpha_A \omega_\alpha\ ,
\end{equation}
of the canonical class of $B$ and the 7-brane classes in \eqref{eq:Kodaira}.

There are several different ways in which the correspondence given by
(\ref{eq:Green-Schwarz-match}) can be derived and/or confirmed.  The
anomaly cancellation conditions provide a consistency check on this
identification; the intersection products between the vectors $a, b_A$ are
given in terms the matter content of the theory through the anomaly
equations, and match with the intersection products between $K, S_A$
in the F-theory geometry \cite{Grassi-Morrison, KMT, KMT-II}.  When
$K, S_A$ span the entire cohomology lattice $H_2 (B,\Z)$ then this
correspondence suffices to prove (\ref{eq:Green-Schwarz-match}).  The
correspondence (\ref{eq:Green-Schwarz-match}) can  also be confirmed
directly from the dual M-theory picture; details of this computation
are given in   \cite{BonettiGrimm}.
In cases where the F-theory model has a heterotic dual, it is furthermore
possible to directly derive the coefficients in the Green-Schwarz term by dimensional reduction of the heterotic 10D theory
and to confirm (\ref{eq:Green-Schwarz-match}) \cite{Honecker}; we describe this connection to 6D heterotic theories in Section \ref{sec:6D-heterotic}.

A fourth approach to deriving the correspondence
(\ref{eq:Green-Schwarz-match})  arises from the expansion of
the curvature-corrected Chern-Simons action of the 7-branes.  
This approach, originally taken by Sadov \cite{Sadov},
is somewhat heuristic as the perturbative 7-brane action is
extrapolated to the nonperturbative regime.  This argument is,
however, the easiest approach to generalize to the analogous 4D
context, so we focus on this method here.
The Dirac-Born-Infeld world-volume action of the 7-branes \cite{DBI}
(reviewed in \cite{Blumenhagen:2006ci})
contains Chern-Simons type couplings that can be written in the schematic form
\begin{equation}
 \int_{M_6 \times S_A}  C_4 \wedge \left(\tr (\hat F^A)^2 -\frac{1}{48}\tr \hat R^2\right)  \,,
\label{eq:7-brane-couplings}
\end{equation}
where $\hat F^A$ is the 7-brane field strength, $\hat R$ is the
curvature two-form restricted to the 7-brane world volume, and $S_A$
are the divisors in $B$ wrapped by the 7-branes.  After dimensional
reduction to 6D, each stack of branes on a divisor $S_A$ associated
with a nonabelian gauge group factor produces the term of the form $B
\cdot S_A\ \tr (F^A)^2$ in (\ref{GS-term}).  All 7-branes, including
those that do not carry nonabelian gauge group factors, should in
principle carry $R^2$ terms.  From the Kodaira condition
\eq{eq:Kodaira}, the sum over these branes gives precisely the class
$[-12K]$, so that the sum over all branes of the $R^2$ terms
reproduces the term of the form $B \cdot K/4\ \tr R^2$ in
\eqref{GS-term}.  This derivation can be understood clearly in the
limit of F-theory discussed by Sen \cite{Sen:1996vd,Sen:1997gv} where
the 7-branes not carrying nonabelian gauge groups combine into
orientifold planes.  We give a more detailed description of the
analogous analysis in the 4D case in Section \ref{sec:4D}.

We see then from the correspondence (\ref{eq:Green-Schwarz-match})
that the canonical class of the base and the divisors carrying the
nonabelian gauge group factors can be read off directly from the
topological couplings in the supergravity action.  To understand this
relationship better it is helpful to discuss the inner product
structure on $SO(1, T)$ vectors somewhat further.  As discussed above,
the inner product $\Omega_{\alpha \beta}$ has signature $(1, T)$.  For
convenience we use a shorthand notation
\begin{equation}
 x \cdot y = x^\alpha \Omega_{\alpha \beta}y^\beta \,.
% \label{eq:}
\end{equation}
The vectors $a, b_A$ are associated with charges of dyonic strings
given by gravitational and gauge theory instantons.  These vectors lie
in an integral lattice.  The integrality of the inner products $a
\cdot a, a \cdot b_A, b_A \cdot b_B$ follows simply from the absence
of anomalies in any 6D supergravity theory, independent of
consideration of quantized string charges \cite{KMT-II}.  Furthermore,
in any consistent theory these vectors must lie in a signature $(1,
T)$ lattice $\Gamma$ that is self-dual (unimodular)
\cite{Seiberg-Taylor}.
In a theory with an F-theory realization, this lattice corresponds to
the second cohomology lattice of the F-theory base
\begin{equation}
\Gamma = H_2 ( B,\Z) \,.
% \label{eq:}
\end{equation}
The intersection product on this lattice corresponds to the inner
product given by $\Omega_{\alpha \beta}$ in the supergravity
theory.
Furthermore, any charge $ x \in \Gamma$ with $j \cdot x > 0$ for all
$j$ in the K\"ahler cone
corresponds to an effective divisor in $B$.
Thus, knowledge of the spectrum of charged string excitations in the
theory provides a complete picture of the cohomology and effective
divisors (Mori cone)
of $B$.  The lattice spanned by $a, b_A$ is in general a sublattice of
the full lattice $\Gamma$.

\subsubsection{Kinetic terms}

We next consider the 
first two terms in (\ref{6Daction}).  These two terms are related by
supersymmetry to the terms in (\ref{GS-term}) \cite{Sagnotti}.
Consider first the
kinetic terms of the 6D vectors with field
strengths $F^A$. 
Independent of the supersymmetry relating these terms to the
topological $BF^2$ terms,
one can
compare the general form of the kinetic term of the 6D vectors with
the kinetic term of the vectors arising in an F-theory reduction. 
This
is done either by an M-theory lift as in \cite{BonettiGrimm}, or by a
direct evaluation of the Dirac-Born-Infeld action for D7-branes.
In the latter route,
by analogy to \eqref{eq:7-brane-couplings} this term is given by
\begin{equation} \label{Fkin6D} 
 \int_{M_6 \times S_A} \text{Tr}\, (\hat F^A \wedge *_8 \hat F^A) 
   = \int_{M_6 } \text{Tr}\, (F^A \wedge * F^A) \cdot
    \frac{\int_{S_A} J_{\rm b}}{\cV_{\rm b}^{1/2}}\ , 
\end{equation} 
where $*_8$ is the
Hodge-star on the 7-brane world-volume, and $J_{\rm b}$ is the K\"ahler form
of the base $B$.  Note that the factor of the base volume $\cV_{\rm b} =
\frac12 \int_B J_{\rm b} \wedge J_{\rm b}$ in \eqref{Fkin6D} 
arises from the Weyl
rescaling of the metric to bring the Einstein-Hilbert term to standard
form.\footnote{One has to perform the rescaling of the 6D metric
  $g_{\mu \nu} \rightarrow \cV_{\rm b}^{1/2} g_{\mu \nu}$.} Expanding the
base K\"ahler form as $J_{\rm b}= v_{\rm b}^\alpha \omega_\alpha$, and comparing
\eqref{Fkin6D} with \eqref{6Daction} one infers 
\begin{equation} 
  j^\alpha =
  \frac{v_{\rm b}^\alpha}{(2\cV_{\rm b})^{1/2}}\ , \qquad \quad j^\alpha \Omega_{\alpha
  \beta} j^\beta = 1\ , 
\end{equation} 
where the latter condition is
automatically satisfied.
Similarly, one can in principle evaluate higher curvature terms in the
Dirac-Born-Infeld action of a D7-brane to fix the 
first term in
\eqref{6Daction}.  In contrast to the kinetic terms of the vectors
$F^A$ this term contains the contraction of the form $K^\alpha
\Omega_{\alpha \beta} j^\beta$, where $K^\alpha$ is
canonical class of the base as in \eqref{Kxi-expansion}.

Finally, we discuss the kinetic term of the two-forms $B_2^\alpha$
that is the remaining term in \eqref{6Daction}. It contains the
metric $G_{\alpha \beta}$, which due to supersymmetry can be given as a
simple expression in terms of the real scalars $j^\alpha$. 
The main purpose of
including this term is  to contrast it with its four-dimensional
analogue (in Section \ref{sec:4D}) where such strong supersymmetry
constraints do not apply.  One notes, however, that $G_{\alpha \beta}$
also admits a small $j^\alpha$ expansion that is valid for large
two-cycle volumes $v^\alpha_{\rm b}$ in the base $B$. Explicitly one finds
\begin{equation} \label{G-exp_6D}
   G_{\alpha \beta} = - \Omega_{\alpha \beta } + \cO(j^2)\ ,
\end{equation}
as discussed in more detail in \cite{BonettiGrimm}.
Hence, in this large-volume limit the kinetic term of the
$B^\alpha_2$ allows us to infer the
intersection matrix $\Omega_{\alpha \beta}$ from the low-energy
effective action.
In six dimensions, this matrix is always equivalent under a linear
field redefinition to the matrix diag($+1, -1, -1, \ldots$); in four
dimensions, however, the analogous structure is more complex.

We close our discussion by noting that in 6D one can in many cases use
the discrete data $T$ and the anomaly lattice to uniquely identify the
F-theory base and topological data of the discriminant locus from
the data of the low-energy theory \cite{KMT, KMT-II}.  When augmented
with information about the dyonic string lattice of the low-energy
theory this data is always sufficient to uniquely determine the topology of the
F-theory base, including the precise structure of effective divisors,
{\it i.e.}~the Mori cone.

\subsection{Examples of 6D F-theory models}
\label{sec:6D-examples}

We give a few brief examples of 6D F-theory models to illustrate some
of the points just reviewed.

\subsubsection{$T = 0$}

The simplest F-theory base for a 6D supergravity model is $\P^2$, with
$\ho = 1$ so $T = 0$.  6D supergravity theories with $T = 0$ were
analyzed extensively in \cite{0} from the point of view of
supergravity constraints, and in \cite{Morrison-Taylor-matter, Braun}
from the point of view of F-theory.  In all $T = 0$ models, $\Gamma
=\Z$, $-a = 3$, since $K = -3H$ where $H$ is the hyperplane generating
$H_2 (\P^2,\Z)$ with $H \cdot H = 1$, and $b_A$ is an integer for each
gauge group where $S_A = b_A H$.

\subsubsection{$T = 1$}

The F-theory bases with $T = 1$ are the Hirzebruch surfaces $\F_m, m
\leq 12$ \cite{Morrison-Vafa-II}.  These are $\P^1$ bundles over
$\P^1$.  A basis for $\htb$ for $\F_m$ is $\Sigma, F$, with $\Sigma$ a section
and $F$ a fiber, and intersection numbers $\Sigma \cdot \Sigma = -m, \Sigma \cdot F =
1, F \cdot F = 0$.  
The irreducible effective divisors in this basis are $\Sigma, F,$
and $q \Sigma + pF$ with $q > 0, p \geq m\,q$.
The generic Weierstrass model over $\F_m$ for $m=
0, 1, 2$ has no gauge group or matter, and for $m = 3, 4, 5, 6, 8, 12$
has a gauge group $SU(3), SO(8), F_4, E_6, E_7, E_8$ with no charged
matter.  We focus here on the structure of the Green-Schwarz terms for
these models.  In the following section these terms are related to the
dual heterotic picture.

There is a natural linear basis for $H^2 (\F_m,\Z)$ given by 
\begin{equation} \label{def-omegafb}
  \omega_{f} = [\Sigma + (m/2) F], \qquad \omega_{b}= [F]\ ,
\end{equation}
where the brackets indicate that we consider the Poincar\'e dual two-forms.
In this basis the inner product is given by
\begin{equation}
 \Omega_{\alpha \beta}=\int_{B_2} \omega_\alpha \wedge \omega_\beta \ ,\qquad \quad \Omega = \left(\begin{array}{cc}
0 & 1\\
1 & 0\\
\end{array} \right) \,.
\label{eq:6D-inner}
\end{equation}
While this basis is not an integral basis for the lattice for $m$ odd,
it will be useful in matching to the heterotic theory.

The 6D two-forms in Type IIB on $\F_m$
arise from the $C_4$ R-R field via the
decomposition $C_4 = B_f \wedge \omega_f + B_b \wedge \omega_b$ into
the two-forms $\omega_f, \omega_b$ introduced in \eqref{def-omegafb}. 
To evaluate the 6D Green-Schwarz terms for $B_f$ and $B_b$
we first determine $a$ for this geometry. 
In the $\omega_f,\omega_b$ basis we have 
\begin{equation} \label{KFm}
   -[K] =  [2 \Sigma + (2 + m)F] = 2 \omega_f + 2 \omega_b \ , \qquad a = (K^\alpha) = (-2,-2)\ ,
\end{equation}  
where $[K]$ is the canonical class of $B_2$. The vector $b$ is determined 
by the wrapping of the 7-brane. For a gauge group factor wrapped on a divisor $S = p\,\Sigma + q\, F$,
one has 
\begin{equation} \label{SFm}
   [S] = p\, \omega_f + (q-p\, m/2) \omega_b \ , \qquad b=(C^\alpha)=(p,q-p\, m/2)\ .
\end{equation}
The Green-Schwarz terms are then obtained by inserting \eqref{KFm}, \eqref{SFm}, and \eqref{eq:6D-inner} 
into the general expession \eqref{GS-term} such that
\begin{equation}
S^{(6)}_{\rm GS} 
= \frac12 \int (B_f  + B_b) \wedge \tr R^2 
 - (p\,B_b +  (q-p\,m/2)B_f) \wedge \frac{2}{\lambda} \tr\ F^2\,.
\label{eq:GS6D_Fm}
\end{equation}

\subsection{Six-dimensional heterotic models}
\label{sec:6D-heterotic}

The $T = 1$ 6D models discussed above are also well understood in a
dual heterotic picture~\cite{Morrison-Vafa-II}.  
Generic Weierstrass
models over $\F_m$ correspond to heterotic $E_8 \times E_8$
compactifications on K3 with $12 \pm m$ instantons in each $E_8$
factor, or for $\F_4$ to heterotic $SO(32)$ compactification on K3.
The 6D Green-Schwarz terms for these theories can be derived directly
from the heterotic 10D action.  This can be seen on the one hand as a
method for confirming the form of the Green-Schwarz terms.  On the
other hand, this can be seen as a simple way of determining the
F-theory dual of the heterotic theories: by finding the low-energy
data associated with a given heterotic model and constructing from
this the F-theory data we can directly determine the F-theory dual of
a given heterotic theory.  This discussion is intended as a warmup for
the 4D case discussed in the following section, where similar
statements hold.

\subsubsection{Heterotic in 10D}

We first recall the 10D heterotic supergravity action with 
gauge groups $SO(32)$ and $E_8 \times E_8$. Since we want to 
determine a 6D action of the form \eqref{6Daction} with the 
duality constraint \eqref{self-dual_tensors} imposed on tensors, it will 
be convenient to  start with a pseudo action in 10D.
This action depends on the heterotic B-field $\hat B$ 
and its dual six-form field $\hat B_6$. 
Throughout this section we use hats (as in $\hat{B}$) to denote 10D
quantities; fields without hats refer to 6D quantities.
A well-known global constraint on 6D heterotic compactifications
arises from the Bianchi identity of the modified heterotic three-form
field strength $\hat H$.  Due to the Chern-Simons connections in $\hat
H$ it satisfies \begin{equation} \label{het_Bianchi} d \hat H = \frac{2}{\lambda}
\Tr \hat F^2- \tr \hat R^2 \ , \qquad \hat H = d \hat B +
\frac{2}{\lambda} w^B_{\rm CS}(F) - w_{\rm CS}(R) \ .  \end{equation} where
$\lambda = 2$ for $SO(32)$, and $\lambda =60$ for $E_8 \times E_8$.
We use the
10D string-frame pseudo-action \footnote{We have used in this action and in \eqref{het_Bianchi}
a normalization of the B-field convenient for heterotic/F-theory duality discussed below.}
\begin{equation} \label{10Daction}
  S^{(10)}_B = \int - \frac{1}{4} e^{-2\phi}\,\hat H  \wedge * \hat H - \frac{1}{4} e^{2\phi}\,\hat H_7  \wedge * \hat H_7  - \frac{1}{3} \hat B \wedge \hat X_8 -  \frac{1}{2} \hat B_6 \wedge \hat X_4\ .
\end{equation}
with the duality $*\hat H=e^{2\phi} \hat H_7$ imposed on the level of the equations of motion. 
Note that the equations of motion of \eqref{10Daction} supplemented by the duality constraint precisely reproduce
the equations of motion of the heterotic action.
The last two terms in \eqref{10Daction} are the 10D Green-Schwarz terms with 
\bea \label{eq:def-X8}
\hat X_8 & = &  
\frac{1}{24} \Tr \hat{F}^4
-\frac{1}{7200}  (\Tr \hat{F}^2)^2
-\frac{1}{240}  \Tr \hat{F}^2\, \tr \hat{R}^2 
+\frac{1}{8}  (\tr \hat{R}^4)
+\frac{1}{32}  (\tr \hat{R}^2)^2 \, ,\\
\hat X_4 &=&  \frac{2}{\lambda}\tr\   \hat{F}^2 - \tr \hat{R}^2\ , 
\eea
where again $\lambda = 2$ for $SO(32)$, and $\lambda =60$ for $E_8 \times E_8$. 
Note that the equations of motion of $\hat B_6$ determined from \eqref{10Daction} 
are precisely the Bianchi identity \eqref{het_Bianchi} via the duality of
the field stregths.

\subsubsection{Green-Schwarz terms and duality to F-theory}

We now describe in detail the derivation of the Green-Schwarz terms
for 6D heterotic compactifications and  the connection through duality
to F-theory.  As we describe in the following section, a very similar
analysis holds in four dimensions.
Green-Schwarz anomaly cancellation in 6D heterotic compactifications
on K3 was first analyzed in \cite{gsw}.  The derivation
of the 6D Green-Schwarz terms from the heterotic theory was done by
Honecker in \cite{Honecker}, and the determination of the structure of
the 6D terms from anomalies was worked out by Erler in \cite{Erler}.
The trace factors needed for this computation are given in
\cite{Erler}.  
Note that the conventions of \cite{Erler, KMT, KMT-II}
differ from those of \cite{gsw, Honecker}.  We follow
the former conventions here.

Consider the heterotic theory compactified on K3, described as a
$T^2$ fibration over $\P^1$.  There is one 6D tensor that we shall
call $B_0$ coming from the 10D B-field $\hat B$ in non-compact directions. 
This $B_0$ is not chiral but a linear combination of 
the self-dual and anti-self-dual tensors that which are
part of the 6D gravity and  tensor multiplets respectively. 
In the action formulation \eqref{10Daction}, with duality condition $*\hat H = e^{2\phi} \hat H_7$
imposed on the level of the equations of motion, one also gets a second 6D tensor from $\hat B_6$.
We denote this 6D tensor arising from $\hat B_6$ wrapped on the 
wrapped on the K3 by $B_1$.  The 10D duality of the $B$ field to $B_6$ reduces to the 
6D duality of $B_0, B_1$ with the inner
product matrix (\ref{eq:6D-inner}). 
The contribution of $B_1$ to the
6D action comes the last term in \eqref{10Daction} and yields
\begin{equation}
 S^{(6)}_{GS}(B_1)=  \frac{1}{2}\int B_1 \wedge (\tr R^2 - \frac{2}{\lambda}\tr\   F^2) \,.
% \label{eq:}
\end{equation}
The contribution of $B_0$ to the 6D action comes from the dimensional
reduction of the 10D Green-Schwarz term involving $\hat X_8$.
To get the 6D action we replace half of the indices in $\hat X_8$ with
internal (compact) indices; we denote curvatures in the compact
directions by $\cR,\cF$.  

In 6D compactifications on $K3$ the Bianchi identity
(\ref{het_Bianchi})
implies \footnote{Here we have fixed the normalization of $\cF,\cR$ such that 
$(2/\lambda) \int_{\rm K3}\tr\cF^2,  \int_{\rm K3}\tr\cR^2$ are integers.}
\begin{equation}
\frac{2}{\lambda} \int_{\rm K3}\tr\cF^2 =
\int_{\rm K3}\tr\cR^2  = 24 \,.
% \label{eq:}
\end{equation}
This implies a fixed total instanton number for $SO(32)$, but allows
for the distribution $12 \pm k$ of instantons between the two gauge
group factors
in the $E_8 \times E_8$ case.

We now consider separately the heterotic $SO(32)$ and $E_8 \times E_8$
theories.  For the $SO(32)$ theory in a generic instanton background,
we replace \cite{Erler}
\begin{equation}
   \Tr \hat{F}^2 = 30\tr \hat{F}^2\ , \qquad \quad \Tr \hat{F}^4 = 24\;\tr \hat{F}^4 + 3 (\tr
\hat{F}^2)^2\ .
\end{equation}  
The $\tr \hat{R}^4,\tr \hat{F}^4$ terms in the fundamental
representation from (\ref{eq:def-X8})
do not contribute in 6D since the curvature in the compact directions
is associated with different indices from the 6D curvatures.  Thus,
(\ref{eq:def-X8}) gives
\begin{equation}
\hat X_{8}^{SO(32)} = \int_{\rm K3} \Big(
- \frac{1}{8} \tr F^2 \, \tr \cR^2
 - \frac{1}{8} \tr R^2 \, \tr \cF^2
+\frac{1}{16}  \tr R^2 \, \tr \cR^2 \Big) \,.
% \label{eq:}
\end{equation}
We thus have the Green-Schwarz couplings
\begin{equation}
S^{(6)}_{SO(32)} = \frac{1}{2}\int (B_1 + B_0 ) \wedge \tr R^2 - ( B_1 - 2 B_0 )\wedge \tr F^2\ .
 \label{eq:gs32}
\end{equation}
Comparing this Green-Schwarz term with the general expression \eqref{GS-term}, we read off 
in the basis $(B_0,B_1)$ with intersection product \eqref{eq:6D-inner} the vectors
\begin{equation} \label{ab-SO32}
  a=(-2,-2)\ , \qquad \quad b = (1, -2)\ ,
\end{equation}
Note that the first entry of $a$ and $b$ is easy to infer by comparing 
the modified heterotic field strength 
\eqref{het_Bianchi} with the general 6D expression \eqref{H3-correct}.
These agree with the F-theory picture \eqref{eq:GS6D_Fm} under the identifications 
\begin{equation} \label{map_B}
  (B_f, B_b ) \ \leftrightarrow  \ (B_0 , B_1)\ .  
\end{equation} 
The vector $a$ in \eqref{ab-SO32} agrees with the F-theory expression
\eqref{KFm}.  In the $SO(32)$ case the vector $b = (1, -2)$ shows that
in F-theory the compactification manifold must be $\F_4$
and the remaining gauge group ($SO(8)$ when maximally broken)
must arise from a 7-brane wrapping the divisor $\Sigma$ on $\F_4$. 
To see this, we use the fact that the vector $b =(1,
-2)$ only encodes an irreducible effective divisor on $\F_m$ for $\F_4$.  For $m
< 4$ the corresponding divisor is not effective and for $m > 4$ it is
not irreducible.  This reproduces the standard picture of
heterotic/F-theory duality in this case \cite{Morrison-Vafa-II}

For  the $E_8 \times E_8$ case, we have a similar analysis.  
Now there
are $12 \pm k$ instantons in the two $E_8$ factors.  
Using the $E_8$ trace normalization and relation \cite{Erler}
\begin{equation}
\Tr \hat{F}^2  =  \tr \hat{F}^2\ , \qquad \Tr \hat{F}^4  =  \frac{1}{100}  \tr \hat{F}^4
\end{equation}
and inserting into (\ref{eq:def-X8}) we get
\begin{eqnarray}
\hat X_{8}^{E_8}  & = &  \int_{K3}  \left(
 \frac{1}{3600} \left[ 
2 \tr F_1^2  \, \tr \cF_1^2 
+ 2 \tr F_2^2 \, \tr \cF_2^2 
- \tr F_2^2 \, \tr \cF_1^2
- \tr F_1^2 \, \tr \cF_2^2 \right]  \right. 
\\
& &\hspace*{0.3in} 
 \left.- \frac{1}{240} \ \left[\tr F_1^2 \, \tr \cR^2  + \tr R^2 \, \tr \cF_1^2 + \tr F_2^2\, \tr \cR^2
 + \tr R^2 \, \tr \cF_2^2 \right] \right. \nonumber \\
&& \hspace*{0.3in} \left.
+\frac{1}{16}  \ \  \left[ \tr R^2\, \tr \cR^2 \right] \right) \,. \nonumber
\end{eqnarray}
Inserting $\int \tr \cR^2= 24, \int\tr\cF_{1, 2}^2 = 30 (12 \pm k)$ gives
the
Green-Schwarz terms
\begin{equation}
 S^{(6)}_{E_8} = \frac{1}{2}\int (B_1 +B_0) \wedge \tr R^2 -
\frac{1}{30} \Big(B_1-\frac{k}{2} B_0 \Big) \wedge \tr F_1^2 -
\frac{1}{30}\Big(B_1 +\frac{k}{2} B_0 \Big) \wedge \tr F_2^2\ ,
% \label{eq:}
\end{equation}
Comparing this with the general 6D expression \eqref{GS-term} gives
\begin{equation}
   a = (-2,-2) \ , \qquad b_1 = (1,-k/2) \ , \qquad b_2 = (1,k/2)\ .
\end{equation}
This is in agreement with the F-theory picture with the 
identification \eqref{map_B} where the remaining components
of the two $E_8$ groups wrap $\Sigma, \Sigma + kF$ on $\F_k$.  
We see that the heterotic/F-theory correspondence of these terms
immediately determines the space for the F-theory dual of each choice
of instanton distribution in the heterotic theory as well as the
locus on which the branes carrying the two gauge group factors are wrapped.
For generic instanton configurations, only one of these gauge groups
remains unbroken; by convention this is taken to be the gauge group
associated with the divisor $\Sigma$ on $\F_k$.

We thus see that by computing the Green-Schwarz terms on the heterotic
side, we can immediately determine the topology and divisor classes of
the base manifold and 7-branes carrying gauge groups for a dual
F-theory model.  Note that on the heterotic side, it possible to have
a K3 that is not elliptically fibered.  In this case there is no clear
F-theory dual.  The determination of the F-theory dual through the
Green-Schwarz terms is only topological, however.  Because the
non-elliptically fibered K3's are in the same moduli space as
elliptically fibered K3 surfaces, they can be reached by a continuous
deformation from models admitting F-theory duals.  It would be
interesting to understand better how this works in the dual F-theory
picture.

\section{Four-dimensional supergravity theories and F-theory vacua}
\label{sec:4D}

We now carry out a similar analysis for 4D supergravity theories and
F-theory constructions.  This section is structured in a parallel
fashion to the 6D story in the previous section, though some of the
technical and conceptual aspects are more complicated.  As in 6D, the
supergravity spectrum and topological terms correspond closely to the
topological structure of 4D F-theory vacua, at least for large-volume
compactifications where the moduli can be clearly identified from the
low-energy theory.  Section \ref{sec:4D-SUGRA} contains some simple
observations on the connection of 4D spectra with F-theory geometry.
We describe the general structure of axion--curvature squared terms in
the 4D action in Section \ref{sec:4D-terms}.  The topological nature
of these terms in 4D encodes much of the relevant structure of the
F-theory compactification geometry, just as the Green-Schwarz $BF^2$
and $B R^2$ terms in 6D encode key aspects of the topology of the
corresponding elliptically fibered F-theory threefold.  This story is
complicated in four dimensions, however, by the appearance of
similar terms associated with additional axion fields, for example 
at weak string coupling from the 10D axiodilaton.  In Section \ref{sec:4D-examples} 
we describe as
examples F-theory compactifications on bases that are complex
threefolds with the structure of a $\P^1$ fibration.  These are dual
to 4D heterotic compactifications over elliptically fibered
threefolds; we describe these models in Section \ref{sec:4D-heterotic}
and show how the axionic--curvature squared terms can be derived from
the heterotic theory and used to identify the topology of the F-theory
dual.

Note that while in six-dimensional supergravity theories the spectrum
of the theory is massless, and the structures visible from F-theory
geometry are clearly apparent throughout the moduli space, the story
is more complicated in four dimensions.  Perturbative and
nonperturbative effects, including the fluxes needed for D3-brane
tadpole cancellation, lift some moduli of the F-theory geometry.
Structures in the action and constraints that are apparent in the
large volume F-theory limit are not protected against perturbative and
nonperturbative corrections, and may be lost or modified in the full off-shell
configuration space of the theory.  In discussing the spectrum and
terms in the 4D supergravity action, we are working in a limit where
the compactification volume is large, and where the spectrum of light
fields, while possibly lifted by fluxes, is still related to the
geometry of the F-theory compactification.  As we discuss at the end
of the paper, going beyond this limit and exploring the implication of
these structures and constraints on the broader off-shell
configuration space is an interesting open problem for further
research.

\subsection{F-theory vacua and 4D spectra}
\label{sec:4D-SUGRA}

We begin by summarizing the field content of 4D ${\cal N} = 1$
supergravity theories, and describing the spectrum that will appear in
any F-theory compactification.  Much of this correspondence is known
\cite{Denef:2008wq,Grimm-F-theory}, but we add some further observations here.
The spectrum of a general $\cN=1$ theory contains a single gravity
multiplet and a number of chiral multiplets $C$, as well as a number
of vector multiplets $V$.  A chiral ${\cal N} = 1$ multiplet contains
a single complex scalar comprising one real scalar and one real
pseudoscalar degree of freedom \cite{Weinberg}
while a vector multiplet contains a
vector as bosonic components.  The standard form of the $\cN=1$
supergravity action is well-known and can be found, 
{\it e.g.}~in
\cite{Wess-Bagger}.

\subsubsection{Scalar spectrum and couplings} \label{moduli_sec}

In an F-theory construction of a 4D ${\cal N} = 1$ theory, we have a
Calabi-Yau fourfold $X$ that is elliptically fibered over a complex
threefold base $\cB_3$.  The origin of the various fields in the 4D
theory is described in \cite{Denef:2008wq,Grimm-F-theory} from the point of view of
F-theory as a limit of M-theory. 
We focus here on neutral scalar fields; charged fields are discussed
in later sections. 
Since a non-Abelian gauge group on the 7-branes 
generically renders the fourfold $X$ singular, 
as in 6D we
use the resolved fourfold $\hat X$
to determine the spectrum of the theory.
In the F-theory picture, there are
$h^{3, 1} (\hat X)$ neutral chiral
multiplets associated with the complex
structure moduli $z^k$ of $\hat X$, or equivalently with the physical
moduli in the
Weierstrass model describing the 7-brane configuration. 
Counting   $h^{3, 1} (\hat X)$
corresponds to considering deformations that preserve the 7-brane gauge group
singularities
that are smoothed
in the resolution from $X$ to $\hat X$. Other chiral
multiplets arise from a basis $\omega_\alpha$ of $H^{1,1} (\cB_3)$ when
expanding the K\"ahler
form $J_{\rm b}$  of $\cB_3$ and R-R 4-form $C_4$ as
\begin{equation} \label{J-exp}
  J_{\rm b} = v_{\rm b}^\alpha \omega_\alpha \ , \qquad C_4 = B_2^\alpha \wedge
\omega_\alpha\ .
\end{equation}
In 4D we can dualize the resulting two-forms to give axions
$\rho_\alpha$ that complexify the fields from the K\"ahler class into
complex moduli $T_\alpha$.  In contrast to the 6D compactifications
discussed in the previous section, there is yet another class of
chiral multiplets, associated to the third non-trivial Hodge number of
a Calabi-Yau fourfold.  In general, there will be $h^{2, 1} (\hat X) -h^{2,
  1} (\cB_3)$ multiplets of this type, which arise in the M-theory picture 
as complex scalars in the expansion of the three-form potential into 
the respective three-form basis of $X$. At weak string coupling these 
fields correspond to modes of the Type IIB R-R and NS-NS two-forms
and the Wilson line modes on the 7-branes.
We decompose the chiral multiplets
into different types, where the numbers of the different types of
fields are related to the Hodge numbers of the F-theory
compactification through 
\begin{eqnarray}
 \label{spectrum-split} 
\cs &=&
h^{3,1}(\hat X) -1\ ,\\ 
\csa &=& h^{1,1}(\cB_3)+1 \ , \\ 
\cp & = &
h^{2, 1} (\hat X) -h^{2, 1} (\cB_3) \ .  
\end{eqnarray}
While the total number of complex scalars is $C = \cs+ C_{\rm
  sa} + \cp$, these different types of scalars have distinct
properties and couplings in the large-volume F-theory limit.  
In particular, the scalars $\csa+ \cp$ are distinguished from the scalars
$\cs$ by the fact that the $\csa+\cp$ can immediately be identified as
having pseudoscalar components with an 
axionic shift symmetry.
At weak string coupling this 
is apparent from the fact that the K\"ahler moduli, the B-field moduli and the dilaton 
are complexified by real scalars arising in the expansion of the Ramond-Ramond forms 
with discrete shift symmetries \cite{Grimm:2004uq}.\footnote{This is equally true 
for the 7-brane Wilson line moduli also contained in $\cp$ \cite{Jockers:2004yj}.} We have therefore grouped the scalar 
containing the degree of freedom of the
axiodilaton with $\csa$ rather than with $\cs$. 

To further distinguish the types of scalar fields, we can study their couplings in the 
effective action. 
As we discuss in more detail in Section \ref{sec:4D-terms}, the $\csa$
scalars generically contribute to axion-curvature squared terms of the
form $\rho R \wedge R, \rho F \wedge F$, where $\rho$ is the
pseudoscalar component.  It is less clear, however, whether the $\cp$
scalars have couplings of this form.  The absence of such couplings
for these axions might be linked with the fact that the scalars $\cp$
have an additional discrete symmetry, as we discuss next.
We recall that at large volume the
definition of the real part of the $h^{1,1}(\cB_3)$ K\"ahler moduli
$T_\alpha$ contains divisor volumes in the base $\cB_3$. The imaginary
part of $T_\alpha$ are the axionic scalars dual to the two-form fields
obtained from reducing $C_4$.  One observes that the $\cp$ complex
scalars $N^a$ appear quadratically in $T_\alpha$, with a coupling
function $d_{\alpha ab}(z,\bar z)$ determined by a holomorphic
functions of the complex structure moduli $z^k$ of $X$.  
Using the corresponding M-theory reduction \cite{Haack:1999zv,Haack:2001jz} one
explicitly finds \cite{Grimm-F-theory}
\begin{equation} \label{Talpha-definition} 
  T_\alpha = \tfrac12
  \kappa_{\alpha \beta \gamma} v^\beta_{\rm b} v^{\gamma}_{\rm b} +
  \tfrac14 d_{\alpha ab}\, (N+\bar N)^a (N+\bar N)^b+ i
  \rho_\alpha\ , 
\end{equation} 
where $v_{\rm b}^\alpha$ are the base
two-cycle volumes introduced in the expansion \eqref{J-exp}, and $\kappa_{\alpha \beta \gamma}$ is the 
triple intersection number on $\cB_3$.  
The leading classical K\"ahler potential $K$ determining the kinetic terms of the scalars 
is given as a function of the base volume $\cV_b$ and the $h^{3,1}(\hat X)$ complex structure moduli of $\hat X$.
It must be evaluated as a function of the complex moduli $T_\alpha,N^a$
and $z^k$ by solving \eqref{Talpha-definition} for $v_{\rm b}^\alpha$ 
and inserting the result into $\cV_b \propto \kappa_{\alpha \beta \gamma}v^\alpha_{\rm b}v^\beta_{\rm b}v^\gamma_{\rm b}$.  
This implies that $K$ is only a function of $T_\alpha +\bar T_\alpha$ and $N^a+\bar N^a$.
Given these expressions we note that the kinetic terms of the 
action have the classical shift symmetries
\begin{equation} \label{shift-symmetry}
  N^a \ \rightarrow \ N^a + i \Lambda^a\ , \qquad T_\alpha \ \rightarrow \ T_\alpha + i  \Lambda_\alpha\ .
\end{equation} 
We expect that generally these symmetries 
will be broken to discrete shifts by quantum states coupling to $T_\alpha,N^a$. 
Further, observe that one has the symmetry $\pi:N^a \rightarrow -
N^a$ due to the quadratic appearance 
of $N^a$ in \eqref{Talpha-definition}. It is tempting to conjecture that this symmetry $\pi$ is preserved 
at the quantum level, and allows one to distinguish the $\cp$ scalars $N^a$ from the others. Such 
a symmetry can also potentially forbid curvature-squared couplings.  

The main structure that we focus on in this section is the
axion-curvature squared terms mentioned above that couple the
pseudoscalar components of the fields $\csa$ to the gravitational and
gauge curvatures.  The existence of such couplings is connected with
the set of quantum string states in the theory that are magnetically
charged under the axion fields.  In general, each axion field obeys a
discrete shift symmetry where the shift of the field lies in the
lattice of possible axionic string charges.  
Note that, just as in six dimensions, the lattice of quantized string
states should arise in four dimensions in any quantum theory of gravity containing
axions under which the strings can carry charges, independent of the
UV completion of the theory.
Each of the fields
of the type $\csa$ contains  a pseudoscalar axion with such a shift symmetry as its imaginary
part, and in the next section we compute the couplings of these axions
to curvature squared terms in the action.  While chiral multiplets in
general include pseudoscalar components as their imaginary
parts,
which may also act as axions under which string
excitations of the 4D theory are magnetically charged, it is less
clear how this works for the other types of scalar fields $\cs, \cp$.
There is no apparent axionic shift symmetry for generic complex
structure moduli $h^{3, 1} (\hat X)$ in the F-theory construction using 
an elliptic fourfold. Nevertheless one can find couplings of the 
scalars $\cs$ to certain $U(1)$-curvature squared terms as 
we discuss in \eqref{fA_tau_T}. It would be very interesting to 
investigate the set of couplings for these scalars in more detail. 
It is possible that away from special limits in the F-theory 
complex structure moduli space, such as the weak string coupling, or the heterotic 
limit, all scalars $h^{3,1}(\hat X)$ mix with other moduli
and correct the curvature-squared 
couplings. For curvature-squared couplings involving the 7-brane 
field strength this was also found in \cite{Grimm:2012rg}.
Indeed, in the context
of mirror symmetry for
${\cal N} = 2$ theories \cite{Hori:2003ic}, complex structure and K\"ahler
moduli are related through a duality symmetry, suggesting that
generically both types of moduli may admit shift symmetries and engage in
couplings to curvature-squared terms.

\subsubsection{Vector spectrum and gauge kinetic functions}

Vector fields in the 4D theory come from two sources. The first class
of vector fields arises in complete analogy to the 6D compactifications. The
nonabelian gauge symmetries arise from the codimension one singularities
of the elliptic fibration of $X$
over the base $\cB_3$. Physically these singularities signal 
the presence of space-time filling 7-branes. 
The rank of the gauge group can be determined by
resolving $X$ to $\hat X$. The total rank $r_{\rm v} $ of the gauge group is
\begin{equation}
  r_{\rm v} = h^{1,1} (\hat X) -h^{1, 1} (\cB_3)-1
\end{equation}
as in \eqref{eq:6D-rank}.  As discussed in section
\ref{sec:6D-basics} this general expression also
counts massless $U(1)$ factors. These are obtained when the elliptic
fibration $X$ has more than one section. 

Let us review the form of the gauge coupling function 
for a non-Abelian gauge group on a stack of 7-branes wrapped 
on divisors $S_A$. 
This coupling can be computed by using an M-theory dual description
\cite{Grimm-F-theory}, or from the 7-brane action at weak coupling as
discussed below, and 
is given at leading order by 
\begin{equation} \label{leading_gauge}
  f_{A} = \frac{1}{2} C_A^\alpha T_\alpha \ , \qquad [S_A] = C^\alpha_A \omega_\alpha\ ,
\end{equation}
where $\omega_\alpha$ is a basis of two-forms of $\cB_3$, and the $T_\alpha$ have been given in \eqref{Talpha-definition}.
This expression for $f_A$ is well-known for D7-branes
\cite{Jockers:2004yj}. 
From the weak coupling 
analysis, however, one expects additional classical corrections to $f_A$. These can 
be induced by fluxes, or by a non-trivial curvature on the brane as we discuss below.
In the F-theory context such corrections have not been studied in full detail. This is due to the 
fact that their M-theory origin is more involved, as recently shown in \cite{Grimm:2012rg}.

In contrast to 6D compactifications one finds in addition
$h^{2,1} (\cB_3)$ $U(1)$ vector fields that arise from expanding
$C_4$ into harmonic three-forms of the base $\cB_3$.
The rank of this abelian part is denoted by
\begin{equation}
  \rp = h^{2,1}(\cB_3)\ ,
\end{equation}
which is equal to the number of such $U(1)$ factors. 
The gauge coupling functions for these $\rp$ vectors 
are given at leading order by \cite{Grimm-F-theory} 
\begin{equation} \label{fA_tau_T} 
 \tau_{\kappa \lambda}(z) =
  \frac{i}{2} \Big(\int_B \beta^\mu \wedge \bar \psi^\kappa\Big)^{-1}
  \int_B \alpha_\lambda \wedge \bar \psi^\mu\ .
\end{equation} 
Here $(\alpha_\kappa,\beta^\kappa)$ is a real symplectic
basis on $\cB_3$, while $\psi^\kappa$ is a basis of $(2,1)$ forms on $\cB_3$
varying with the complex structure moduli $z^k$. In other words, at
this leading order
$\tau_{\kappa \lambda}$ only depends on the complex structure moduli
$z^k$. The imaginary part of $\tau_{\kappa \lambda}$ thus couples 
to $F^\kappa \wedge F^\lambda$ inducing a coupling of type axion-curvature 
squared to the $h^{2,1}(\cB_3)$ Ramond-Ramond $U(1)$ vectors. Similarly one 
expects subleading corrections to \eqref{leading_gauge} depending on 
the complex structure moduli $z^k$.

\subsubsection{Fluxes, D3-brane tadpole and chiral spectrum}

A key difference from the situation for 6D F-theory compactifications is the fact 
that 4D vacua allow for a non-trivial background flux. 
Including such fluxes in four-dimensional F-theory constructions is
the subject of substantial current work \cite{Denef:2008wq}-\cite{Krause:2012yh}.
In fact, such fluxes are often necessary for tadpole cancellation
and have to be present in a consistent vacuum. This leads to an intriguing
interplay of geometric data and flux data. It will be a far reaching task to 
unify both into a common framework. Here we make some basic observations 
that will be  useful in the analysis below of the 4D effective action.

To begin with, we note that there are three types of background fluxes
in F-theory: R-R and NS-NS three-form flux in the bulk, and two-form
fluxes on the 7-branes.  While an individual description of these
fluxes can be 
difficult to integrate with the Weierstrass description of an F-theory model,
there is a natural lift of these fluxes into a single
type of four-form flux $G_4$
that can be interpreted as an actual
four-form on a smooth geometry $\hat X$
in the dual M-theory compactification, where $G_4$ is the field strength
of the M-theory three-form.  A general $G_4$ induces a 4D
superpotential as well as a D-term.  The superpotential is given by
$W(z) = \int G_4 \wedge \Omega$ \cite{Gukov:1999ya}, and depends
holomorphically on the $h^{3,1}(\hat X)$ complex structure moduli of
$\hat X$.  The large volume D-term depends on the K\"ahler moduli via
the K\"ahler form on $\hat X$.  It will be useful to introduce the
the matrix
\begin{equation}
    \Theta_{\Sigma \Lambda} = \int_{\hat X} \omega_\Sigma \wedge \omega_\Lambda \wedge G_4\ .
\end{equation}
where $\omega_\Sigma$ is a basis of two-forms of $H^2(\hat X,\mathbb{Z})$ on the resolved fourfold, 
including all new classes $\omega_{i_B}$ obtained after resolution of gauge group $G_B$ 
singularities for non-Abelian 7-branes. 
Components of $\Theta_{\Sigma \Lambda}$ determine the D-terms.
The D-terms arise from gaugings of the shift symmetries \eqref{shift-symmetry} of 
the imaginary part of the scalars $T_\alpha$ given in \eqref{Talpha-definition}. In the M-theory dual Coulomb branch 
description the gauge-invariant derivative is given by 
\beq
   DT_\alpha  = dT_\alpha + i \Theta_{\alpha i_B} A^{i_B}
\eeq
where ${i_B}$ labels the forms $\omega_{i_B}$ arising from 
resolving the gauge group singularities for $G_B$. In the 4D F-theory 
compactification one has to replace $\Theta_{\alpha i_B}$ with an
adjoint valued matrix $\underline{\Theta}_{\alpha B}$ and the invariant derivative 
takes the form 
\beq
    DT_\alpha  = dT_\alpha + i \Tr(\underline{\Theta}_{\alpha B}\, A^{B})
\eeq
Note that $\underline{\Theta}_{\alpha B}$ corresponds to a non-Abelian flux background on the $B$th 7-branes 
and thus will break the gauge group $G_B$.

Let us stress here, that both the superpotential 
as well as the D-terms give mass to some of the $\cs+\csa$ moduli. This complicates the identification 
of light states in the 4D effective theory. Since, however, the masses both from the superpotential 
and the D-terms are suppressed by a volume factor of higher power than for the masses of the KK-modes,
one can identify light fields at large volume.

It is crucial to note that $G_4$ has to obey various constraints. 
First, it has to be quantized appropriately, since $G_4 + c_2(\hat X)/2$ has
to be an integral class \cite{Witten:1996md}. This condition 
has recently been analyzed systematically for F-theory geometries in \cite{Collinucci:2010gz,Collinucci:2012as}.
Secondly, certain components of
$\Theta_{\Sigma \Lambda}$ 
have to vanish in order that $G_4$ lifts to an F-theory flux
and preserves 4D Poincar\'e invariance \cite{Denef:2008wq,Marsano:2011hv,Grimm:2011sk}. 
Fluxes are crucial to induce a 4D chiral matter spectrum as recently 
studied in \cite{Donagi-Wijnholt,Beasley-hv,Beasley-hv2}, \cite{Braun:2011zm}-\cite{Krause:2012yh}.
In general the components  
\beq  \label{res_Theta}
    \Theta_{i_A j_B}= \int_{\hat X} \omega_{i_A} \wedge \omega_{j_B} \wedge G_4\ ,
\eeq 
with $\omega_{i_A},\omega_{j_B}$ resolving the gauge group singularities for $G_A,G_B$ on 
one or more 7-branes, can be non-zero.  Physically these components of $\Theta_{\Sigma \Lambda}$ 
carry the information about the 4D chiral matter spectrum integrated out at one loop in the M-theory 
compactification \cite{Grimm:2011fx}. 
  
In 4D compactifications the fluxes and geometry 
are linked via the well-known global consistency condition ensuring cancellation of
3-brane tadpoles. 
In the M-theory language, this tadpole constraint
is 
\begin{equation} \label{D3-tadpole}
  \frac{\chi(\hat X)}{24} = \frac12 \int G_4 \wedge G_4 + N_{3}\ ,
\end{equation}
where $\chi(\hat X)$ is the Euler character of the resolved Calabi-Yau fourfold. 
Here $N_3$ is the number of 3-branes, which are point-like objects in $\cB_3$. 
The number of independent components of $G_4$ depends upon the Hodge
number $h^{2, 2} (X)$.  A linear relation between the Hodge numbers on
a Calabi-Yau fourfold
\cite{Sethi:1996es, Klemm-lry}
\begin{equation}
h^{2, 2} (X)= 44 + 4h^{1, 1} (X)+2h^{2, 1} (X)-4h^{3, 1} (X)
\label{eq:22-constraint}
\end{equation}
shows that the Euler character can then be written as
\bea
\chi (\hat X) &=& 6 (8 + h^{1, 1} (\hat X)+h^{3, 1} (\hat X)-h^{2, 1} (\hat X))  \nonumber \\
        &=& 6 (9 + \csa + \cs + r_{\rm v} -(\cp
+ \rp)) \,.
\label{eq:general_euler}
\eea
This links the light spectrum with the fluxes via \eqref{D3-tadpole}.
Note, however, that fluxes also make it hard to identify massless 
moduli since, as mentioned above, the flux-induced D-term will generate a potential 
for the fields introduced in section \ref{moduli_sec}.
In general,
there will be many discrete choices of flux associated with a given
geometric F-theory background.  This can be either achieved by using
$G_4$ fluxes or introducing a number $N_3$ of 3-branes.

Note that the presence of D3-branes leads to additional light 
degrees of freedom corresponding to the moduli of the D3-branes. For $N_3$ separated D3-branes they 
shift the number of fields
\begin{equation} \label{D3-shift}
  \Delta C_{\rm cs} = 3 N_3 \ , \qquad \Delta r_{\rm v} = N_3\ ,
\end{equation}
where the scalars arise from the three complex positions of the D3-brane in $\cB_3$.
Note that each D3-brane comes with an additional four-dimensional 
$U(1)$ gauge symmetry which is generically unbroken. This leads to a link \eqref{D3-shift} 
of the number of $U(1)$'s with the number of massless deformations. This should 
be contrasted with the case of 7-branes, or even D7-branes, where the brane $U(1)$'s can be massive at the Kaluza-Klein 
scale due to a geometric St\"uckelberg term \cite{Jockers:2004yj,Grimm:2011tb}. This is true both in 4D and 6D since in both cases 
the 7-brane embedding into the base $\cB_3$ can have a non-trivial topology. We do not work with theories here that  have separate D3-brane degrees
of freedom.

\subsection{Couplings in the 4D supergravity action from F-theory}
\label{sec:4D-terms}

In this section we discuss some specific terms in the action of the 4D
$\cN=1$ effective supergravity theories that arise by compactifying
F-theory on the elliptically fibered fourfold $X$.  As in the 6D
discussion \eqref{6Daction} we focus on terms in the action that
contain information about the topological data of the compactification
manifold $X$.  It is important to remember that
$4D,\cN=1$ is much less protected against corrections than its 6D
counterpart due to the smaller number of supersymmetries and
space-time dimensions.  It is therefore more challenging to extract
the geometric data from a low-energy 4D supergravity theory than in
6D.  As discussed above, this connection can be made most clearly in
the large-volume limit.  We find it useful in this discussion to frame
part of the analysis involving scalar-axion fields in
terms of a dual 4D picture where these fields are analogs of 6D tensor
multiplets.

\subsubsection{Topological couplings in the F-theory effective action}
\label{sec:4D-couplings}

We now wish to focus  on particular terms in the $\cN=1$ effective
action involving the scalar-axion fields $\csa$.
In particular, we consider terms in the theory
that are analogous to the couplings
(\ref{GS-term}), and have the form
\begin{equation}
  S^{(4)}_{ax} = \frac{1}{8}\int \frac{1}{2} a^{\cA}  \rho_{\cA}\ \tr R \wedge R
+ \frac{2}{\lambda_A} b_A^\cA \rho_\cA \ \tr F^A \wedge F^A \,.
 \label{eq:4D-topological}
\end{equation}
In these couplings, $\rho_\cA$ are the axions appearing as the
imaginary parts of the $\csa$ scalar fields.
The index $\cA$ is taken to be $0$ for the axion associated with the
axiodilaton at weak coupling, and $\alpha=
1, \ldots, h^{1, 1} (\cB_3)$ are the remaining $\csa$ fields.
As in 6D, $a^\alpha$
describes the canonical class of the F-theory base, while $b_A^\alpha$
contains information about the divisor classes on which the nonabelian
gauge group factors are wrapped.  
Just as in six dimensions, the terms (\ref{eq:4D-topological}) can
play a role in anomaly cancellation through the generalized
Green-Schwarz mechanism \cite{Green:1984sg,Blumenhagen:2005mu,Plauschinn}.
These terms appear, however, independent of the need for anomaly
cancellation, and are present even in theories without unbroken gauge groups.
The appearance  in these terms of the canonical
class of the F-theory base and the classes carrying the 7-branes
associated with gauge groups plays a key role in the
correspondence between the low-energy theory and the F-theory
construction, just as in six dimensions. 
The 4D story is more complicated,
however, due to the existence of extra axions that
do not tie directly into the divisor geometry in F-theory. For example, at weak string coupling the dimensionally
reduced 10D axiodilaton will appear in the curvature-squared couplings, despite the fact that 
it does not admit a two-form interpretation in 6D.
We  discuss such additional axions in section \ref{additional_axion}.
As we  describe in more
detail below in section \ref{sec:triple_intersect}, the couplings
\eqref{eq:4D-topological} admit a dual description when the
axions are dualized to two-forms and lead to corrected field strengths
of the form \eqref{H3-correct}. 

We now describe how the terms appearing in the action
\eqref{eq:4D-topological} are determined in an F-theory reduction on a
Calabi-Yau fourfold.  We concentrate first on the terms involving the
$h^{1, 1} (\cB_3)$ axions in $T_\alpha$. To determine $b_A^\alpha$ we note
that the $F^2$ coupling is given by the imaginary part of the gauge
coupling function $f_A$.  For a 7-brane these can be extracted
at weak string coupling using an argument analogous to
\eqref{eq:7-brane-couplings}, as we show below. Alternatively, as
discussed above, one can use the duality
between M-theory and F-theory to derive $f_A$
\cite{Grimm-F-theory}. 
Either approach gives a leading gauge
coupling function \eqref{leading_gauge}  that is linear in $T_\alpha$.
Comparing this with \eqref{eq:4D-topological} gives
\begin{equation}
  b^\alpha_A = C^\alpha_A \ , \qquad \quad [S_A] = C^\alpha_A
  \omega_\alpha\ ,
\label{eq:identification-b}
\end{equation}
where $S_A$ are the divisors in $\cB_3$ wrapped by the 7-branes.
The weak-coupling 7-brane analysis below also describes
the higher curvature terms $R^2$ with coupling $a^\alpha$.
The upshot is that, just as in 6D, $a^\alpha$ corresponds to the canonical class of the
F-theory base manifold  through
\begin{equation} \label{identification_a_alpha}
      a^\alpha = K^\alpha\ , \qquad \quad c_1(\cB_3)= - K^\alpha \omega_\alpha\ . 
\end{equation}

One expects that the expression for $a^\alpha$ can also be determined via
an M-theory reduction as in the 6D/5D reduction
\cite{BonettiGrimm}. This is more involved, however, than in the 6D
case, due to the fact that the $ R \wedge R$ term couples to the axion
$\rho_\alpha$ rather then the two-form $B_2^\alpha$  that is the 4D
dual of $\rho_\alpha$. 

We now give the weak coupling 7-brane analysis showing that the
identifications (\ref{eq:identification-b}) and
\eqref{identification_a_alpha} are indeed correct, and fix the
numerical factors in the latter coupling.  This analysis proceeds in
analogy to the similar analysis for 6D compactifications.  We consider the Chern-Simons
couplings of the D7-branes  and O7-planes. The branes will
admit $SU(N^{\hat A}_{\rm D7})$ gauge groups, such that $\lambda_{\hat A}=1$ in \eqref{eq:4D-topological}. 
We focus here particularly on the
terms coupling to the R-R four-form $C_4$ which admits the expansion
$C_4 = \rho_\alpha \tilde \omega^\alpha$, where $\tilde \omega^\alpha$
are the four-forms on $\cB_3$ dual to $\omega_\alpha$.  Inserting this
expansion into the Chern-Simons actions and integrating
over the compact directions one obtains \bea \label{CS_actions}
S^{CS}_{\rm D7}( R,F) &=&- \frac{1}{2}
\int_{M^{3,1}} C^\alpha_{\hat A} \rho_\alpha \Big( \frac{1}{96}
\text{tr}( R \wedge R) N_{\rm D7}^{\hat A} + \frac{1}{2}
\text{tr}(F^{\hat A} \wedge F^{\hat A}) \Big) + \ldots \ , \nonumber
\\ S^{CS}_{\rm O7}( R) &=& + 2
\int_{M^{3,1}} \tilde C^\alpha \rho_\alpha \Big(- \frac{1}{192}\,
\text{tr}( R \wedge R) \Big)+ \ldots\ , \eea where $C_{\hat A}^\alpha$
is the restriction to the $\hat A$th D7-brane stack, and $N_{\rm
  D7}^{\hat A}$ is the number of D7-branes on the $\hat A$th
stack. Here $\hat A$ will run only over indices labeling the
D7-branes.  The restriction to the O7-plane is denoted by $\tilde
C^\alpha$. Note that in F-theory, both $C_{\hat A}^\alpha$ and $\tilde
C^\alpha$ are combined into $C_A^\alpha$. The fact that in F-theory
the O7-plane is split into two 7-branes is captured by the relative
factor of $2$ when comparing the tr$( R^2)$ for the D7-brane and
O7-planes after using $\mu_{\rm O7} = -4 \mu_{D7}$. 
From the $F \wedge F$ terms we confirm (\ref{eq:identification-b}).
We are now also in a
position to show \eqref{identification_a_alpha}. At weak coupling  the
discriminant in F-theory splits through
\begin{equation}
  -12 [K] =  [\Delta] =2 [S_{\rm O7} ] + N^{\hat A}_{\rm D7}  [S_{\hat A}]  =  (2 \tilde C^\alpha + N^{\hat A}_{\rm D7} C^\alpha_{\hat A} ) \omega_\alpha \ ,
\end{equation}
where the first equality is the Kodaira constraint \eqref{eq:Kodaira}. Hence, one infers $12 K^\alpha = - (2 \tilde C^\alpha + N^{\hat A}_{\rm D7} C^\alpha_{\hat A} )$, which inserted 
into the sum of the actions \eqref{CS_actions} yields the
identification \eqref{identification_a_alpha}. 
Note that we have determined the numerical 
factor in $a^\alpha$ by comparison to the result \eqref{fA_tau_T} for $f_A$.

\subsubsection{An additional axion} \label{additional_axion}

As discussed above, we have included an additional scalar-axion field
contributing to $\csa$ thereby treating it on a similar footing to the $h^{1,1}(\cB_3)$
scalars $T_\alpha$; we denote this field by $T_0$. This can be motivated by 
the fact that there are two limits in the complex structure moduli space of $\hat X$, where
one scalar-axion field is singled out. 

Firstly, at weak string-coupling $T_0$ corresponds to the dimensionally reduced 
axiodilaton $\tau_{\rm IIB}$ setting
\begin{equation}
T_0 = - i \tau_{\rm IIB} = e^{-\phi} - i C_0\ .
 \label{eq:t0}
\end{equation}
In this weak coupling limit Re$\, T_0$ is large, which is the analog to the Re$\,T_\alpha$
in the large-volume limit of $\cB_3$. Using an expansion of the Chern-Simons actions 
of D7-branes and O7-branes one can derive the couplings $a^0,b^0_A $ of Im$\, T_0$ to $R^2$ and $(F^A)^2$.
However, away from the weak coupling limit in a generic F-theory compactification 
the axiodilaton $\tau_{\rm II\cB_3}$ is no longer a well-defined 4D field, since it admits 
$SL(2,\mathbb{Z})$ monodromies around the 7-branes. The appropriate coordinates 
are now the $h^{3,1}(\hat X)$ complex structure deformations that
contain this weak-coupling degree of freedom. We thus expect couplings 
of the form 
\beq
   \frac{1}{2}\, \tilde a(z)\, \tr R \wedge R + \frac{2}{\lambda_A}\, \tilde b_A(z)\, \tr F^A \wedge F^A\ , 
\eeq
where $ \tilde{a}(z),\tilde{b}_A(z)$ admit appropriate monodromy properties for a 
given Calabi-Yau geometry. At weak coupling the functions $\tilde{a},\tilde{b}_A$ can be 
expanded into a term linear 
in Im\,$T_0$ with exponentially suppressed corrections.  We do not have a clear derivation of these
terms in a general F-theory setup.  In principle it should be
possible to compute these couplings by gluing together contributions 
from all 7-brane actions focusing on the axiodilaton coupling. Note that the described situation is 
very similar to the complex structure dependent couplings \eqref{fA_tau_T} for the Ramond-Ramond $U(1)$ vectors, and hence 
to the analogous $\cN=2$ story for Calabi-Yau threefold compactifications. $ \tilde{a}(z),\tilde{b}_A(z)$
are expected to be complicated functions of the complex structure moduli that depend on 
the point in moduli space and have some characteristic expansion with coefficients determined by 
the topological data of the base $\cB_3$ and gauge bundles on the 7-branes.

A second limit in which the couplings $ \tilde{a}(z),\tilde{b}_A(z)$ can be expanded into 
a term linear in a single axion with exponentially suppressed corrections can be accessed for F-theory 
geometries with a heterotic dual. We will make this precise in Section \ref{sec:4D-heterotic},
and derive these couplings through the duality to heterotic theory.

Note that, as mentioned in the introduction to this section, giving up 
the the  large-volume limit for $\cB_3$ one expects to also lose the linearity 
in the $T_\alpha$, and all axions may mix non-trivially.  It is not
clear from the F-theory point of view how these couplings can be computed,
however.  We leave further investigation of this question to future work.

\subsubsection{Remarks on the triple intersection numbers and
  identification of F-theory geometry}
\label{sec:triple_intersect}

We have seen so far how the Hodge numbers, canonical class, and
7-brane divisors of an F-theory compactification are encoded in the
corresponding 4D supergravity theory.  As in 6D, a key part of the
topological structure of the F-theory compactification geometry lies
in the intersection form of the F-theory base.  In 4D, this
intersection form is the triple intersection product $\kappa_{\alpha \beta
  \gamma}$.  In contrast to the 6D story, however, the 
triple intersection product is not immediately visible
in a general 4D $\cN=1$ supergravity theory.
Due to perturbative and non-perturbative corrections, the kinetic terms of the 
scalars are not protected.
Only at leading order in the large-volume limit in an 
F-theory compactification is the triple product visible.  

The situation 
is relatively clear when the scalars $T_\alpha$ can be replaced by linear multiplets containing 
as bosonic components a real scalar $L^\alpha$, which is the Legendre
dual $\text{Re}\ T_\alpha$ with respect to the K\"ahler potential, 
and a two-form $B^\alpha$ dual to $\text{Im}\ T_\alpha$ \cite{Cecotti:1987nw,Binetruy:2000zx}. This can be done if $\text{Im}\ T_\alpha$ possesses a shift symmetry \eqref{shift-symmetry}.
Then, as in 6D in \eqref{H3-correct}, the field-strength of $B^\alpha$ is given by 
\begin{equation} \label{H3-correct-4D}
   H_3^\alpha = d B_2^\alpha + \frac12 a^\alpha w_{\rm CS}(R)+2 \frac{b^\alpha_A}{\lambda_A} w_{\rm CS}^A(F)\ , 
\end{equation}
For small $L^\alpha$ one can then expand the metric $ \tilde G_{\alpha \beta} $
for $L^\alpha$ and $H_3^\alpha$ as
\begin{equation} \label{expand_tildeG}
   \tilde G_{\alpha \beta}  \propto \kappa_{\alpha \beta \gamma} L^\gamma + \ldots\ ,
\end{equation}
which is the analogue to \eqref{G-exp_6D} in the 6D action.  At large
volume the 4D couplings then contain the triple intersections
$\kappa_{\alpha \beta \gamma}$ with the third index contracted with
$L^\alpha$. It remains to be shown how much of this structure survives quantum 
corrections. In particular, away from the large-volume limit where $L^\alpha$ can be large, the structure \eqref{expand_tildeG} 
is not expected to be preserved. 
Nonetheless, in the large-volume limit the triple intersection
coefficients are contained in the leading order term in the expansion (\ref{expand_tildeG}).

We have thus outlined a way in which much of the topological structure
of the F-theory base $\cB_3$ and fourfold $X$ can be identified from
the 4D supergravity action, at least in the large-volume limit where
the spectrum remains light and corrections are small.  We have
identified the Hodge numbers, canonical class, and intersection form
of the F-theory base.  From Wall's theorem \cite{Wall}, the homotopy
type of a compact complex 3-manifold can be identified from the Hodge
numbers, triple intersection form, and first Pontryagin class $p_1
(\cB_3)$. In principle, information about $p_1$
will be contained in higher-order terms in the string
action. In particular, at weak string coupling one 
can use the higher-curvature corrections in the D7-brane 
Chern-Simons action to find couplings of the axiodilaton
to the $p_1$ restricted to the branes. A more complete 
understanding of the allowed couplings away from special 
limits in the F-theory complex structure field space might
thus yield the desired information. 
In principle then,
enough data to reconstruct the topology of the F-theory
compactification space is contained in the
structure of the 4D supergravity theory at large volume.  
While providing additional information, the couplings of additional axions can 
also complicate the reconstuction of the geometry. 
To reconstruct the
topology of the F-theory base it is necessary to know which axions
correspond to the complex structure moduli.  While in some cases this
is clear in the large-volume limit, in general this may require
further information.  
Knowing the topology of the F-theory base is also not sufficient to
completely determine the geometry.  It is also necessary to know the
complex structure on the base to fully identify the theory.  As in 6D,
the quantum spectrum of supersymmetric charged string solitons encodes
the Mori cone of the threefold base in a 4D compactification; this is
discussed further in Section \ref{sec:4D-lattice}.
In
the specific examples we discuss below for heterotic/F-theory duality,
the data given in the 4D theory, coupled with an identification of
axions, is sufficient to uniquely determine the F-theory geometry.

\subsection{4D F-theory examples with $\P^1$ fibered base}
\label{sec:4D-examples}

We now discuss a general class of examples of F-theory
compactifications, where the base $\cB_3$ of the Calabi-Yau fourfold
is a $\P^1$ fibration over some complex surface $\cB_2$.  This is the
class of 4D F-theory models that admit a duality to heterotic
compactifications \cite{fmw,Bershadsky-jps,Curio-Donagi,Andreas}.  
We describe the general topological structure of these fibrations in
section \ref{sec:top_of_P1fibrations}, and discuss explicit examples
with $\cB_2 = \P^2$ in section \ref{sec:P1P2examples}.  Duality to the
heterotic string is described in Section \ref{sec:4D-heterotic}

\subsubsection{On the geometry of general $\P^1$ fibrations} \label{sec:top_of_P1fibrations}

We begin by reviewing some generalities regarding $\P^1$ fibered bases
$\cB_3$, following \cite{fmw, Andreas}. We consider $\P^1$ fibrations with a
section $\Sigma$.  Such
fibrations can be characterized by two two-forms $r,t$ that are
obtained as follows.  First, consider the sum $\cO \oplus \cL$ of two
complex line bundles $\cO$ and $\cL$.  The base $\cB_3$ is the
projectivization of  this vector bundle, {\it i.e.}~$\cB_3 = \P(\cO
\oplus \cL)$.  There are two distinguished two-forms $r,t$ on $\cB_3$,
which are given by the first Chern classes
\begin{equation} \label{def-rt}
   r = c_1(\cO(1)) = [\Sigma]\ , \qquad t = c_1(\cL)\ .
\end{equation}
Here $\cO(1)$ is a line bundle on $\cB_3$  that restricts to the typical line bundle of each $\P^1$ fiber of $\cB_3$, and 
$r$ restricts to the $\P^1$ hyperplane class. The classes $r,t$
satisfy 
%\cite{fmw}
\begin{equation} \label{r(t+r)}
  r (r+t)=0\ .
\end{equation} 
In the evaluation of the low-energy couplings 
we need the characterisic classes \footnote{Note that $\int_{\cB_3} c_1(\cB_3) c_2(\cB_3) = 2 \int_{\cB_2} c_1(\cB_2) c_2(\cB_2) = 24$, by using the 
fact that $\chi_0 (\cB_3)=1$.}
\bea \label{Chern-base}
   c_1(\cB_3) &=& c_1(\cB_2) + (2 r + t)\ ,\\
   c_2(\cB_3) &=& c_2(\cB_2) + c_1(\cB_2)(2 r+t)\ , \nonumber
\eea
as can be shown by using the adjuction formulas.

In order to perform the F-theory reduction we introduce the basis of
two-forms $\omega_\alpha = (\omega_f,\omega_i)$ on $\cB_3$, and expand
$C_4 = B_2^\alpha \wedge \omega_\alpha$.  
The four-dimensional
two-forms $B_2^\alpha$ are dual to the axions $\rho_\alpha$.  The
internal two-forms $\omega_i$ are pulled back from two-forms of
$\cB_2$. In summary we introduce the basis
\begin{equation} 
\omega_f = r+ \frac{t}{2}  \ , \qquad    \omega_i   \ . 
\end{equation} 
Using \eqref{r(t+r)} one 
infers the triple intersections
\begin{equation}
   \kappa_{ijk} = 0 \, ,\quad \kappa_{f ij} = \kappa_{ij} \, ,\quad  \kappa_{ff i} = 0\,,   \quad \kappa_{fff} = \tfrac14  t^i t^j \kappa_{ij}\ .
\end{equation}
with intersection form $\kappa_{ij} = \int_{\cB_2} \omega_i \wedge \omega_j$, and $t^i$ appearing in the 
expansion $t = t^i \omega_i$.
This is the higher-dimensional analog of \eqref{def-omegafb}, \eq{eq:6D-inner}.
In this basis we use \eqref{Chern-base} to determine 
the vector $K^\alpha$ as 
\begin{equation}
   (a^\alpha_F) \equiv (K^\alpha) = (-2,K^i)\ ,
\label{eq:4D-a}
\end{equation}
where $K^i$ are the coefficients of the canonical class of $\cB_2$ in
the basis $\omega_i$.  
In general, 7-branes carrying gauge groups can be wrapped on an
arbitrary effective irreducible four-cycle in $\cB_3$ with Poincar\'e-dual 
class
\begin{equation}
[S] = p \omega_f + \sum_{i}q_i \omega_i,
\label{eq:4D-branes}
\end{equation}
similar to the 6D case \eqref{SFm}.

\subsubsection{Examples: $\P^1$ fibrations over $\P^2$} \label{sec:P1P2examples}

A simple example for $\P^1$ fibered base spaces are threefolds 
that are described by $\P^1$ bundles over $\cB_2 = \P^2$.  These manifolds, which we
denote by $\tilde{\F}_k$ are close relatives of the Hirzebruch surfaces
$\F_m$.  They are studied in the physics context, for example, in
\cite{Klemm-lry, Mohri, Berglund-Mayr}.  These
models have a simple toric description\footnote{The toric 1D cones  
for $\tilde \F_k$ are generated by $e_1,e_2,e_3$, $-e_3$ and $-e_1-e_2-k\, e_3$, where $e_i$ are the 
unit vectors of $\mathbb{R}^3$.}, 
but can be equally specified using the construction of section \ref{sec:top_of_P1fibrations}.
We denote by $H$ the hyperplane class of the two-fold base $\P^2$
pulled back to $\cB_3$, and $\Sigma$ the class of a section
corresponding to $r$.
Furthermore, we identify $t$ specifying the $\P^1$ bundle as 
\begin{equation} \label{eq:t-ansatz}
    t = k [H] \ . 
\end{equation}
Clearly, using \eqref{Chern-base} and the fact that $c_1(\P^2) = 3[H]$ and $c_2(\P^2) = 3[H]^2$ we find 
\begin{equation} \label{eq:c1c2-tildeFk}
   c_1(\tilde \F_k) = 2 [\Sigma] + (3+k) [H] \ , \qquad c_2(\tilde \F_k)
   = 6 [\Sigma \cdot H] + (3 + 3k) [H]^2 \ .
\end{equation}
 The triple intersections are simply given by
 \begin{equation}
\Sigma\, H^2 = 1,\quad H^3 = 0, \quad
\Sigma^2 \, H = -k, \quad
\Sigma^3 = k^2, \; \;
 \; \;
\label{eq:tfm-intersections}
\end{equation}
The first condition is inherited from the base $\cB_2 = \P^2$, while the second corresponds 
to the fact that three elements of the base cannot intersect for a fibration.
The last two are a trivial consequence of the general fact that 
$r (r + t) = 0$ for $\tilde \F_k$ implies that $\Sigma^2 = -k \Sigma \cdot H$.

The threefolds $\tilde{\F}_k, k = 0, 1, 2, 3$ are bases for generic
elliptically fibered Weierstrass models without codimension one
singularities that would impose a gauge group on the 4D theory.  For
higher values of $k$, the divisor $\Sigma$ is rigid and $f$, $g$ and
$\Delta$ must vanish to a degree that mandates the appearance of a
nonabelian gauge group over that divisor.  For example, $\tilde \F_4$
carries a minimal gauge group $SU(2)$, and $\tilde \F_{18}$ carries an $E_8$
over $\Sigma$.  $\tilde \F_k$ cannot be a good F-theory base for $k > 18$
\cite{Berglund-Mayr}.  
F-theory compactifications on the threefolds $\tilde \F_k$ can be dual to
heterotic compactifications on Calabi-Yau threefolds that are
elliptically fibered over $\P^2$.  We can use the
analysis of Section \ref{sec:4D-couplings} to read off which divisor
classes in $\tilde \F_k$ must support the gauge group, and hence
identify which of the surfaces $\tilde \F_k$ is needed for the F-theory
dual from knowledge of the
bundle structure on the heterotic side. 
We  discuss the 4D heterotic models next.

\subsection{Four-dimensional heterotic models and heterotic/F-theory duality}
\label{sec:4D-heterotic}

The Green-Schwarz two-form--curvature squared terms in 6D theories of
the form $BF^2, B R^2$ provide an illuminating connection between
heterotic and F-theory models, as discussed in section
\ref{sec:6D-heterotic}.  A similar relationship holds in four
dimensions, which we describe in this section. 
The relationship between 4D heterotic and F-theory
compactifications was studied in detail in the seminal work by
Friedman, Morgan, and Witten \cite{fmw}.  They showed that many
heterotic bundles admit a ``stable degeneration limit'' in which
duality to F-theory can be clearly understood.  This work, and
subsequent developments following \cite{Curio-Donagi} have led to an
extensive study of this duality; for a review of some of this work see
\cite{Andreas}.  As in the 6D heterotic/F-theory duality, the 4D
duality relates the heterotic theory on a Calabi-Yau manifold that is
elliptically fibered over a base $\cB_2$ to an F-theory model on a
$\P^1$ fibration over $\cB_2$.  Friedman, Morgan, and Witten relate
the bundle structure on the heterotic side to the twisting of the
$\P^1$ fibration on the F-theory side for $E_8 \times E_8$ heterotic
theory.  We find here that this identification of bundle structure
with twisting follows naturally from the structure of the
axion--curvature-squared terms in the 4D action, and that in many
cases the F-theory geometry dual to a given heterotic model is
uniquely determined by the structure of these terms.  The locus of the
7-branes carrying the gauge group action on the F-theory side is also
uniquely determined by the axion--curvature-squared terms arising from
a given bundle structure on the heterotic side.  These considerations
are independent of the type of bundle construction, and give a
topological picture of heterotic/F-theory duality that is valid for
the $SO(32)$ theory as well as for the $E_8 \times E_8$ theory.

As described in Section \ref{sec:6D-gs}, for a 6D F-theory
compactification the coefficients of the $B R^2$ term are components
of a vector in the string charge lattice characterizing the canonical
class of the F-theory base, while the coefficients of the $BF^2$ term
characterize the divisor class on which the 7-branes giving each
nonabelian gauge group factor are wrapped.  The corresponding
coefficients can be computed directly in the heterotic theory, as
described in Section \ref{sec:6D-heterotic}, by reduction of the 10D
$H^2$ and Green-Schwarz terms.  For any 6D supergravity that has dual
descriptions in terms of heterotic and F-theory compactifications,
this correspondence in the 6D Green-Schwarz terms provides a direct
topological description of the duality.  In particular,
given any heterotic compactification with an F-theory dual, by
computing the 6D Green-Schwarz terms we can read off the canonical
class of the F-theory base and the
divisor
classes on which the 7-branes are wrapped in the dual F-theory model.
Essentially the same story holds in four dimensions.  

In this section we compute the axion--curvature squared terms of the
form $\rho R^2$ and $\rho F^2$ for a general heterotic
compactification on an elliptically fibered threefold with section.
This not only serves as a check on the structure of these terms as
described for F-theory models, but also provides a direct means for
identifying the structure of the dual F-theory model.
In Section \ref{sec:general-heterotic} we describe the general class
of heterotic models and compute the axion--curvature-squared terms in
this general context.  Section \ref{sec:heterotic-examples} describes
some explicit examples of this approach to understanding
heterotic/F-theory duality.

\subsubsection{General heterotic models with F-theory duals}
\label{sec:general-heterotic}

We consider a general 4D ${\cal N} = 1$ supergravity model that has
both a weakly coupled large-volume heterotic description and a
large-volume F-theory description in appropriate regimes.  On the
heterotic side, the 10D $SO(32)$ or $E_8 \times E_8$ heterotic theory
is compactified on a Calabi-Yau threefold $Z_3$ that has the form of an
elliptic fibration over a base manifold $\cB_2$ that is a complex
surface.  
We assume that the elliptic fibration has one (but not more than one)
section, so that  $h^{1, 1} (Z_3) = h^{1, 1} (\cB_2) + 1$.
On the F-theory side, the compactification manifold is an
elliptic fibration over a complex threefold $\cB_3$ that is a $\P^1$
fibration over $\cB_2$.

The axions on the heterotic side consist of an axion $\chi_0$ arising
from wrapping the 10D six-form $\hat B_6$ (dual to the two-form $\hat B$) on the
compactification space $Z_3$, and $h^{1,1}(Z_3)$ axions $\chi^I= (\chi^i,\chi_f)$ from wrapping
the 10D two-form $\hat B$ on the two-cycles of $Z_3$. One thus expands
\begin{equation}
\hat B =b_2 + \chi^I \omega_I =b_2 + \chi_f \omega_{\cB} + \sum_{i}\chi^i \omega_i \,,
% \label{eq:}
\end{equation}
where $\omega_\cB$ is the two-form Poincar\'e dual to the base $\cB_2$ of $Z_3$, and
$\omega_i$ are dual to the two-cycles in $H_2 (\cB_2)$. The four-dimensional 
two-form $b_2$ is dual to $\chi_0$. 
We can now
follow essentially the same analysis as in 
Section \ref{sec:6D-heterotic} to
determine the 4D axion--curvature squared couplings from this
compactification.

Before doing that,  first note that the global consistency conditions arising 
from the Bianchi identity \eqref{het_Bianchi} now split into $h^{1,1}(Z_3)$ conditions
\begin{equation}
    c_{I} - \lambda_{I} = 0\ , 
\end{equation}
where we have defined
\begin{equation}
c_{I} = \int_{Z_3} \omega_{I} \wedge  \tr \cR^2 =  \int_{Z_3} \omega_{I} \wedge  c_2(Z_3) \ ,\qquad \quad   \lambda_I = \frac{2}{\lambda}  \int_{Z_3} \omega_{I} \wedge  \Tr \cF^2  \ ,
\end{equation}
with $\lambda=2$ for $SO(32)$, and $\lambda=60$ for $E_8 \times E_8$ as above.
For an elliptically fibered threefold $Z_3$ we can determine $c_{I}$ more explicitly, by 
using  \cite{fmw} 
\begin{equation} \label{eq:c2Z3}
c_2 (Z_3) = 11c_1 (\cB_2)^2 + c_2 (\cB_2) + 12 c_1 (\cB_2) \wedge \omega_\cB \,,
\end{equation}
which gives $c_2 (Z_3)$  in terms of classes of $\cB_2$, and we have suppressed
the pullback to $Z_3$. Evaluated on a basis one finds
\bea \label{def-cicf}
    c_{i} &=& - 12 \,\kappa_{ij} K^j\ ,\\ 
    c_{f} &=& \int_{\cB_2} c_2 (Z_3) = \int_{\cB_2} c_2(\cB_2) -c_1(\cB_2)^2 = -8+2 h^{1,1}(\cB_2)\ , \nonumber
\eea
where $\kappa_{ij}$ is the intersection form of the $\omega_i$ on
$\cB_2$, and $K$ is again the canonical class of the base $\cB_2$.

We can now determine the four-dimensional couplings of the axions. 
The coupling of $\chi_0$ to $F^2$ and $R^2$ comes from the 
kinetic term of $\hat H$
in ten dimensions by using $* \hat H = \hat H_7$. It is given by
\begin{equation} 
  \chi_0 \, \Big (\tr\, R^2 - \frac{2}{\lambda}\tr\, F^2 \Big) \, .
% \label{eq:}
\end{equation}
We can then analyze the
contributions from $\hat B \wedge \hat X_8$
separately in the $SO(32)$ and $E_8 \times E_8$ theories as in six
dimensions; the algebra follows in a practically identical fashion in
both cases to the six-dimensional analysis.

For the $SO(32)$ theory, the couplings are
\bea 
  S^{(4)}_{SO(32)}  &=& \frac{1}{2}\int \Big( \chi_0 + \frac{1}{24} \chi^I c_{ I}  \Big) \tr\,R^2 - \Big( \chi_0 -\frac{1}{12} \chi^I c_{I}  \Big) \tr\, F^2 
\label{eq:4D-so}\\
                    &=& \frac{1}{2}\int \Big( \chi_0 - \frac{1}{2} \chi^i \kappa_{ij} K^j  + \frac{1}{24}\chi_f c_f \Big) \tr\,R^2 \nonumber \\
                    &&  \quad  \     -\Big( \chi_0 + \chi^i \kappa_{ij} K^j  - \frac{1}{12}\chi_f c_f \Big) \tr\,F^2\ . \nonumber
\eea
In order to determine the vectors $a$ and $b$ we compare 
\eqref{eq:4D-so} with the general form \eqref{eq:4D-topological}. 
We first define 
\begin{equation} \label{def-tilderho}
  \tilde \chi_0 =  8 \chi_0 \ , \qquad \tilde \chi_i = 8 \kappa_{ij} \chi^j\ , \qquad \tilde \chi_f =  8 \chi_f\ .
\end{equation}
In the basis $( \tilde \chi_0 , \tilde \chi_i , \tilde \chi_f)$ we read off, using 
$\lambda=2$ for $SO(32)$, vectors \footnote{Note that we can formally obtain the 6D result \eqref{ab-SO32} for a twofold base by setting $K_{\P^1} = - 2$ and 
dropping the last entry in \eqref{a-SO32-4D}.}
\begin{equation} \label{a-SO32-4D}
   a = \Big(-2,K^i,- \tfrac{1}{12} c_f \Big)\ , \qquad b = \Big(1,K^{i},- \tfrac{1}{12} c_f \Big)\ ,
\end{equation}
with $c_f = - 8 + 2 h^{1,1}(\cB_2)$ as shown in \eqref{def-cicf}. 

Let us now turn to the discussion of the $E_8 \times E_8$ theory.
In this case we have to specify two bundles $V_1
\oplus V_2$. We introduce the general split of the curvature four-forms
\begin{equation} \label{Fisplit}
\frac{1}{30}\tr\cF_i^2  =  \eta_i \wedge \omega_\cB + \zeta_i\ ,
\end{equation}
where $\eta_i$ is a two-form, and $\zeta_i$ is a four-form inherited from 
$\cB_2$. The Bianchi identity \eqref{het_Bianchi} implies in 
cohomology 
\begin{equation}
   \tr \cF_1^2 +\tr\cF_2^2 = 30\, \tr \cR^2\ .
\end{equation}
Using the second Chern class of $Z_3$ as given in \eqref{eq:c2Z3}
we have
\bea \label{othersplits}
   \eta_1 + \eta_2 &=&12 c_1(\cB_2) \ , \\
  \zeta_1 + \zeta_2 &=&  11c_1 (\cB_2)^2 + c_2 (\cB_2) \equiv  \cC_2  \ .
\eea
To satisfy these two conditions we can make the general Ansatz
\begin{align} \label{def-etachi}
  &\eta_1 = 6c_{1}(\cB_2)+ \tilde t\ ,& \qquad &\eta_2 = 6c_{1}(\cB_2)- \tilde t\ ,&\\
   &\zeta_1= \tfrac12 \cC_2 + \Phi\ ,& \qquad &\zeta_2 =\tfrac12  \cC_2 - \Phi\ .&  \nonumber
\end{align}
where $\tilde t, \Phi$ are a two-form and a four-form inherited from $\cB_2$.

For the $E_8 \times E_8$ theory, the couplings are
\begin{equation} 
  S^{(4)}_{E_8}  = \frac{1}{2}\int \big( \chi_0 + \tfrac{1}{24} \chi^I c_{ I}  \big) \tr\,R^2 - \big( \chi_0 -\tfrac{1}{12} \chi^I \tilde c_{I}  \big) \tr\, F_1^2 
                    - \big( \chi_0 + \tfrac{1}{12} \chi^I \tilde c_{I}  \big) \tr\, F_2^2 \ ,
\label{eq:4D-E8}
\end{equation}
where we can treat the vector of axions $\chi^I = (\chi^i,\chi_f)$.
In this expression we have defined  
\begin{equation}
  \tilde c_I = \int_{Z_3} \omega_I \wedge (\tilde t \wedge \omega_\cB + \Phi)=  (-1)^i \int_{Z_3} \omega_I \wedge \big(\tfrac12 \tr \cR^2-\tfrac{1}{30}\tr \cF_i \big)\ . 
\end{equation} 
where $i=1,2$, and the second identity is a trivial consequence of \eqref{Fisplit}, \eqref{othersplits} and \eqref{def-etachi}.
We can evaluate $\tilde c_I$ for the basis $\omega_i, \omega_\cB$ to find
\begin{equation}
    \tilde c_i = \kappa_{ij} \tilde t^j \ , \qquad \tilde c_f = \int_{\cB_2} \Phi - \tilde t^2 \ .
\end{equation}
Using these expressions and comparing \eqref{eq:4D-E8} with the general action \eqref{eq:4D-topological} and $\lambda=60$, we 
find in the basis $( \tilde \chi_0 , \tilde \chi_i , \tilde \chi_f)$ defined in \eqref{def-tilderho} the vectors
\begin{equation} \label{ab-E8}
   a = \big(-2,K^i,-\tfrac{1}{12}c_f \big) \ ,\qquad \begin{array}{l}
                                   b_1 = \Big(1,- \tfrac{1}{2} \tilde  t^i,- \tfrac{1}{12} \tilde c_f \Big)\\[.2cm] 
                                   b_2 = \Big(1,\tfrac{1}{2} \tilde t^i ,\tfrac{1}{12} \tilde c_f \Big)
                                 \end{array}
\end{equation}

From the results (\ref{eq:4D-so}) and (\ref{eq:4D-E8}) we can 
directly read off topological information about the dual F-theory
model.
The F-theory axions $\rho_0, \rho_\alpha$ must be related 
to the heterotic axions through
\begin{eqnarray}
\rho_0 & \quad \leftrightarrow \quad& \tilde \chi_f \\
\rho_b & \quad \leftrightarrow \quad& \tilde \chi_0\\
\rho_i & \quad \leftrightarrow \quad& \tilde \chi_i \,,
\end{eqnarray}
where $\rho_0$ is the F-theory axion from the 10D IIB axiodilaton,
$\rho_b$ comes from $C_4$ integrated over the base $\cB_2$, and $\rho_i$ come
from $C_4$ integrated over four-cycles in $\cB_3$ obtained from curves
in $\cB_2$.
Comparing the expressions for the vector $a$ from
\eqref{ab-E8} and \eqref{a-SO32-4D} to  (\ref{eq:4D-a}) we see that
this identification of axions gives a clear match between the
canonical class of the bases $\cB_2$ used in the heterotic and
F-theory constructions.

Considering the axion--$F \wedge F$ terms, comparing the vectors $b$
from \eqref{ab-E8} and \eqref{a-SO32-4D} to (\ref{eq:4D-branes}) 
we can directly read off the divisor classes on which the 7-branes
supporting the gauge group factors are wrapped on the F-theory side
from the information about the heterotic bundle, just as in 6D.  From
the $\chi_0$ terms we see that every brane wraps the base cycle $\Sigma$
on the F-theory side precisely once, again as in the 6D story.  While
many F-theory models could be constructed on $\cB_3$ with gauge groups
wrapping other cycles that do not wrap the base once, these models
will have no perturbative heterotic dual and require  the introduction
of heterotic 
five-branes.
For the $SO(32)$ theory the brane also wraps the classes on the base
with twice the multiplicity of the canonical divisor $K$.
For $E_8 \times E_8$ theories, the
divisor classes on which the branes giving the gauge group factors are
wrapped depends upon the class of the difference $\tilde{t}$ appearing in
the splitting of the bundle into two components. 
The fact that these divisors must be effective and irreducible
places constraints on the line bundle $t$ describing the $\P^1$ bundle
over $\cB_2$.  
This leads naturally to the identification
\begin{equation}
  t^i  \quad \leftrightarrow \quad \tilde t^i\  \,, \label{eq:dual-identification}
\end{equation}
corresponding to the association between the heterotic bundle and
F-theory $\P^1$ fibration found in \cite{fmw}.  In some situations
this identification is  the unique possibility that satisfies
the constraint on the divisor classes carrying the gauge group
factors.  We give below a simple class of examples where this
identification is uniquely determined, in analogy with the 6D
heterotic/F-theory duality story.

It is interesting to note that with the identification
\eq{eq:dual-identification}
and the heterotic axion--curvature-squared terms
\eq{a-SO32-4D},
the heterotic $SO(32)$ theory
is always dual to F-theory on the same space that carries the dual to
the $E_8 \times E_8$ theory where the decomposition \eq{def-etachi} is
\begin{equation}
\eta_1 = 8c_1 (\cB_2), \;\;\;\;\;
\eta_2 = 4c_1 (\cB_2) \,.
\label{eq:}
\end{equation}
This matches nicely with the observation that on the F-theory side, in
the weak coupling orientifold limit
described by Sen \cite{Sen:1996vd,Sen:1997gv}, the O7-planes carry
precisely 1/3 of the total Kodaira bound of $-12K$. Thus the orientifold limit
fits naturally with the $SO(32)$ heterotic string, as also suggested by 
the gauge groups and representations that appear in that limit. 

Note that the analysis here does not depend upon knowing anything
about the construction of the specific bundles on the heterotic side,
only on the decomposition of the bundle in a way that satisfies the
Bianchi identity.  Various constructions of bundles are known for the
heterotic theory, including spectral cover \cite{fmw,Bershadsky-jps,
  Curio-Donagi,Andreas-Curio,Andreas} and monad \cite{monad-1} (see
{\it e.g.}  \cite{monad} and references therein for a recent overview)
constructions.  The topological information about the duality found
here should be valid and agree with all of these constructions in the
appropriate limits.

Finally, we turn to the terms of the form $\chi_f\tr\ R^2,
\chi_f\tr\ F^2$ in the heterotic picture.  These correspond to
couplings proportional to $\rho_0$ in the F-theory picture.  
As discussed earlier, we do not have a way to directly compute these
couplings in F-theory.  Heterotic/F-theory duality gives us the answer
for this computation in the weak-coupling large-volume heterotic
limit.  We leave as an open problem the connection of this result to a
more general computation in the F-theory context.

\subsubsection{Heterotic/F-theory duality: examples}
\label{sec:heterotic-examples}

We conclude the heterotic/F-theory discussion with a brief description
of how the general duality dictionary described above applies for the
examples $\tilde \F_k$ introduced in section~\ref{sec:P1P2examples},
and make some general statements about a broader class of examples.

We begin with the $E_8 \times E_8$ heterotic string on a
Calabi-Yau that is elliptically
fibered over $\cB_2 =\P^2$.  From (\ref{othersplits}) the total
instanton number is 36, so we distribute the instantons $18 \pm m$ to
the two bundles, with $\tilde{t} = m[H]$.
The F-theory dual will have a base $\cB_3$ that is a $\P^1$ fibration
over $\P^2$, and thus is the threefold $\tilde{\F}_k$, with $t =
k[H]/2$.  
This matches with the results of \cite{Berglund-Mayr}.
From
(\ref{ab-E8}) we see that the divisors carrying the gauge group in the
F-theory picture are $\Sigma + (k/2 \pm m/2)H$.  These are only
irreducible effective divisor classes if $k = m$.  Thus, the
topological heterotic/F-theory determined by the vectors  $a, b$
controlling the axion-curvature squared terms uniquely determines the
F-theory manifold and gauge brane divisor classes for any given
topological class of bundle on the heterotic side.

A similar story holds for the $SO(32)$ theory as in the $E_8 \times
E_8$ case and as in 6D.  For $SO(32)$, from (\ref{a-SO32-4D}) the
contribution of $b$ in the base is $-3$ since $K = -3 H$, so we
see
that the branes on the F-theory side are wrapped on $\Sigma + (k/2-
3)H$.
It follows that the $\P^1$ bundle on the F-theory side is always
$\tilde{\F}_6$, matching with the fact that the unbroken gauge group on
this base is generically $SO(8)$.

A much broader class of examples of 4D heterotic/F-theory duality can
be found in the recently produced list of 61,539 toric bases $\cB_2$
that can be used for an elliptic fibration with section of a
Calabi-Yau threefold \cite{mt-toric}.  In principle, any of these
spaces can be used to construct an elliptically fibered threefold, and
the dual F-theory model will be on a base $\cB_3$ that is a $\P^1$
fibration over $\cB_2$.  The simplest case to consider is one where
the heterotic theory is an $E_8 \times E_8$ theory with the bundle
evenly split, so $\tilde{t} = 0$.  This will be dual to an F-theory
model on the trivial bundle $\P^1 \times\cB_2$.  In general, the toric
bases in this list have many ``non-Higgsable clusters''
\cite{Morrison-Taylor} giving rise to multiple copies of gauge groups
such as $E_8$ and $F_4$ with no charged matter.  These should correspond
on the heterotic side to singular Calabi-Yau geometries, where
enhanced gauge groups appear at the singular loci.  Understanding the heterotic
moduli that may smooth these singularities on the 
F-theory side, where fluxes may be involved, is an interesting
direction for further work.

\section{Geometrical constraints in 6D and 4D}
\label{sec:constraints}

Not all classical supergravity theories can be realized in F-theory.  The
geometric structure of F-theory places specific constraints on the
spectrum and action of supergravity theories that admit an F-theory
construction.  
In six dimensions,
some geometrical F-theory constraints correspond to anomaly
cancellation conditions or other known consistency conditions for a quantum
low-energy supergravity theory.  In other cases, it is not known
whether the geometrical constraints from F-theory are necessary for
consistency of any quantum 6D supergravity theory.  In this section we
review the structure of these constraints in 6D and describe related
constraints on F-theory models in four dimensions.  Some constraints
on 4D supergravity theories arising from F-theory are closely
analogous to 6D F-theory constraints that can be understood in terms
of anomalies in 6D.  In 4D, however, these constraints cannot be
understood from gravitational anomaly cancellation or other known
consistency conditions.  Other constraints in 4D are analogous to the
F-theory constraints on 6D theories for which there is as yet no
macroscopic explanation in terms of consistency of supergravity
theories; these constraints also lack macroscopic 4D interpretations.
In general, the constraints that we find on 4D theories can only be
clearly formulated in the large-volume F-theory limit where the lifted
moduli of the theory are still light.  Unlike ${\cal N} = 1$ theories
in six dimensions that have 8 supercharges and are quite constrained,
${\cal N} = 1$ theories in four dimensions with only 4 supercharges
are relatively unconstrained.  Away from the large-volume F-theory
limit the fields become massive and couplings mix, and it is difficult
to identify global constraints.  Nonetheless, the F-theory constraints
we describe here may provide a useful window on some aspects of the
general structure of 4D ${\cal N} = 1$ string vacua.

We begin in  Section \ref{6Dgeometrical_constraints} with a
description of the constraints provided by geometry for 6D F-theory
compactifications.  We then describe the analogous 4D structures in
Section \ref{sec:4D-constraints} and discuss the nature and range of
validity of the 4D constraints.
The underlying geometric formulae we use here are not new; in
particular, the connection between the Euler character of an
elliptically fibered Calabi-Yau manifold and the geometry of the base
manifold was described in detail in \cite{Klemm-lry, Andreas-Curio}.
The emphasis here, however, is on framing these geometric relations in
terms of the spectrum and terms in the action of the low-energy
supergravity theory, where they become constraints on theories
admitting an F-theory realization.

\subsection{Geometrical constraints on  6D effective supergravity theories}
\label{6Dgeometrical_constraints}

In Section \ref{sec:6D} we described the correspondence between data
in the spectrum and action of a 6D supergravity theory and topological
aspects of the F-theory compactification geometry giving rise to the
6D supergravity theory.  We now turn to the question of what
constraints are imposed by F-theory on this 6D supergravity data.

We focus here on constraints for the restricted class of theories that
contain no charged matter fields, since these are the simplest
constraints to understand geometrically.  This
restriction still gives us insight into a wide range of F-theory
compactification spaces, since for many (but not all) 6D supergravity
theories, all matter fields can be Higgsed so that the generic model
over the given base has no charged matter
\cite{Morrison-Taylor, mt-toric}.
There is also a deep relationship for 6D theories between F-theory
geometry and anomaly constraints on theories with charged matter
\cite{Sadov, Grassi-Morrison, KMT-II, Morrison-Taylor-matter,
  Grassi-Morrison-2}.  While this structure is substantially richer
than for theories without matter, the description of matter in four
dimensions makes the 4D analogue of these constraints more
complicated, and we do not attempt to systematically understand
constraints on 4D theories with matter in this paper.

\subsubsection{Constraints on theories without gauge groups}

Let us start by considering 6D theories with no gauge group (or where
the gauge group has been completely Higgsed).  In this case the number
of vector fields vanishes, $V = 0$, and there are no vectors $b_A$
appearing in the action.  As  shown in \cite{Morrison-Taylor},
these theories have $T \leq 9$, and correspond to F-theory
compactifications on base surfaces that are 
generalized del Pezzo surfaces, containing no effective irreducible
curves of self-intersection $-3$ or less.
There are two constraints that are imposed
by F-theory on the numbers of scalar and tensor fields $H, T$ in
the spectrum and the gravitational Chern-Simons vector $a$
\begin{eqnarray}
a \cdot a & = &  9-T \label{eq:6D-constraint-1}\\
273 & = &  29T  + H \label{eq:6D-constraint-2}
\end{eqnarray}
These constraints are also quantum consistency conditions for the
supergravity theory based on  gravitational
anomaly cancellation.  Thus, in this case
the F-theory constraints correspond to known consistency conditions on
the low-energy theory.  As we will see, F-theory imposes an analogous
constraint on 4D theories, though in that case the constraint is on
light rather than massless fields and is only clearly formulated in
the large-volume F-theory limit.  There is no known anomaly condition
in the low-energy theory associated with the corresponding 4D
constraint.

\vspace*{0.05in}
\noindent {\bf Derivation of constraints from F-theory}
\vspace*{0.05in}

For completeness, and for comparison with the analogous 4D story,
we now review explicitly how the constraints (\ref{eq:6D-constraint-1}),
(\ref{eq:6D-constraint-2}) follow from the
geometry of F-theory.  Related arguments appear in
\cite{Morrison-Vafa-II, Aspinwall-Katz-Morrison, Grassi-Morrison, Grassi-Morrison-2,Park}.  
In terms of the F-theory geometry, using the
correspondences (\ref{eq:t-h11}) and (\ref{eq:Green-Schwarz-match}),
(\ref{eq:6D-constraint-1}) is the condition
\begin{equation}
K \cdot K = 10 -h^{1, 1} (B) \,.
\label{eq:6D-constraint-1-again}
\end{equation}
This can be proven geometrically as follows:
The holomorphic Euler characteristic of the base is \cite{Griffiths}
\begin{equation}
\chi_0 (B) = \frac{1}{12} \int_B (c_1^2 + c_2)
\label{eq:hec}
\end{equation}
where $c_i$ are the Chern classes of $B$.  The left-hand side of
\eq{eq:hec} is just $\chi_0(B) = 1$ since $h^{0,0} =1 , h^{0,i} = 0,
i>0$.  Hence, $12 = \int c_1^2 +\chi(B)$, where
\begin{equation}
\chi(B) = \int_B c_2 = \sum_i (-1)^i b^i = 2 + h^{1,1} \,,
% \label{eq:}
\end{equation}
with  $b^i$ the Betti numbers of $B$.
It follows that
$10 -h^{1,1}(B) = \int c_1^2= K \cdot K$.

Now consider the constraint (\ref{eq:6D-constraint-2}).  In the
absence of a nonabelian gauge group, the
F-theory compactification is described by a smooth
Calabi-Yau threefold  $X$ with an elliptic fibration over the base $B$.
In order
to completely specify $X$ one has to give the global properties of the
elliptic fiber.  The number of massless
$U(1)$'s in the 6D theory is given by 
\begin{equation} n_{U(1)} = h^{1,1}(X) -
h^{1,1}(B)-1\ , \end{equation} which coincides with the rank $r$ given
in \eqref{eq:6D-rank}.  The number $n_{U(1)}$ can be also determined
by counting the number of sections of the elliptic fibration
(Mordell-Weil group).  With no abelian gauge group, $n_{U(1)}=0$, and
$r=0$ in \eqref{eq:6D-rank}.  For spaces with $n_{U(1)} = 0$, there is a
Weierstrass model of the form \eq{eq:Weierstrass}, and the Euler
character of $X$ is related to the topology of the base by
\cite{Klemm-lry}
\begin{equation}
 \chi (X) = -60 \int_B c_1 (B)^2 = - 60 \, \Omega_{\alpha \beta}
 K^\alpha K^\beta = 2 (T - H+ 3) \,, 
\label{eq:6D-Euler}
\end{equation}
where we have used 
%$\int_B c_1(B)^2 = \Omega_{\alpha \beta} K^\alpha K^\beta $  and
\eqref{generalchi_6D}, and all hypermultiplets are neutral.  Combining
this with (\ref{eq:6D-constraint-1}), we have
\begin{equation}
30 (9-T) = H-T - 3 \; \;
\Rightarrow \; \;
273 = H + 29T \,,
 \label{eq:6D-scalar-constraint}
\end{equation}
and we have shown that the constraint (\ref{eq:6D-constraint-2}) 
follows from F-theory geometry.

\vspace*{0.05in}
\noindent {\bf Transitions between F-theory bases}
\vspace*{0.05in}

Before including nonabelian gauge groups in the discussion, we now
briefly discuss the transitions among vacua with different
numbers of tensor multiplets.  
All F-theory bases for 6D supergravity theories are connected through
transitions associated with blowing down $-1$ curves until a minimal
model surface is reached \cite{Grassi, mt-toric}.
The constraints (\ref{eq:6D-constraint-1}) and
(\ref{eq:6D-constraint-2}) must hold globally on the space of theories
without gauge groups and can be characterized by their invariance
under transitions that preserve this property.

As is well known \cite{Seiberg-Witten, Morrison-Vafa-II}, the
tensionless string transitions in 6D theories associated with blowing
down curves to points in the base $B$ produce a change in field
content of the 6D theory where a tensor field is exchanged for $29$
scalar fields.  From the 6D supergravity point of view, the number of
fields replacing the tensor multiplet must be $29$ to satisfy the
anomaly condition \eq{eq:HVtext}.  There are also several ways in
which this change in field content can be understood from the F-theory
geometry.  Starting from the Weierstrass model on the blown-down base
it can be seen that 29 scalars corresponding to Weierstrass
coefficients must be tuned to blow up a point in the base
\cite{Witten-phase-mf}.  Equivalently, using the expression
\eqref{eq:6D-Euler} for the Euler character of $X$ one can infer this
change from purely geometric arguments.

To extract the change in the Hodge numbers using 
\eqref{eq:6D-Euler} we continue to
restrict to the case of having 
no nonabelian gauge symmetries and no massless 
$U(1)$'s. We consider a modification of the 
base $B$ by blowing up a generic point into an exceptional curve $E$ 
in a new base $B'$, denoting the blow-down map by 
\begin{equation} \label{transition_map_6D}
  \pi: B' \rightarrow B
\end{equation}
The first Chern class changes according to 
\begin{equation} \label{changec1_6D}
    c_1(B') = \pi^* c_1(B) - [E]\ .
\end{equation}
Using the 
fact that $[E] \wedge \pi^* [c_1(B)] = 0$, as well as
$E^2=-1$ we have
\begin{equation}
    \chi(X') = \chi(X) - 60\ . 
\end{equation} 
Since  one new K\"ahler class
is gained
in the blow-up, we infer using
$\chi = 2 (h^{1,1}-h^{2,1})$ that $29$ elements of $h^{2,1}$ are lost.
This implies by \eqref{eq:t-h11} and  \eqref{eq:6D-neutral} that the spectrum changes as
\begin{equation} \label{Hodge_change}
   \Delta T = 1 \ , \qquad \Delta H_{\rm neutral}= - 29\ ,
\end{equation}
Hence,   in the blow-up transition one tensor multiplet 
is added to the spectrum, while $29$ neutral hypermultiplets are lost.

From the point of view of the complete 
Calabi-Yau threefold, this transition corresponds to blowing up a singular point
into a del Pezzo $8$ surface. After performing a flop transition this 
becomes a del Pezzo $9$ (half K3), which is an elliptic fibration over the
exceptional curve $E=\mathbb{P}^1$ in $B'$ \cite{Morrison-Vafa-II}.

\subsubsection{Constraints on theories without charged matter}
\label{sec:6D-without}

We now generalize the discussion further and allow for nonabelian
gauge groups in the 6D effective theory.  This leads us to consider a
broader class of F-theory compactifications on singular elliptic
fibrations.  
We continue to restrict attention to models without charged matter
fields, as discussed above.

We  state briefly some additional constraints that arise
from F-theory geometry for 6D theories without matter.  
For theories with nonabelian gauge groups,  these constraints involve
the vectors $b_A$ associated with each gauge group factor, and the
total number of vector multiplets $V =\sum_A {\rm dim}_{G_A}$.
In theories without charged matter there are no
abelian gauge group factors.
The constraints from F-theory geometry are
\begin{eqnarray}
 ( -12 a-\nu_A  b_A) \cdot  b_A  & = &0  \label{eq:6D-constraint-3}\\
b_A \cdot b_B & = & 0, \;\;\;\;\; A \neq B \label{eq:6D-constraint-4}\\
29T -273 & = & V-H \label{eq:6D-constraint-5}
\end{eqnarray}
Like the constraints (\ref{eq:6D-constraint-1}) and
(\ref{eq:6D-constraint-2}), these F-theory constraints can be
understood in the supergravity theory from anomaly cancellation;
similar constraints in four dimensions, however, have no analogous
understanding in terms of the low-energy theory.  

We now briefly describe the constraints just listed from the point of
view of F-theory geometry.   We begin with
(\ref{eq:6D-constraint-3}), (\ref{eq:6D-constraint-4}).  
Matter fields in F-theory arise  either
from
codimension two singularities in the discriminant locus or from higher
genus topology of divisor classes $S_A$.  Codimension 2
singularities can either occur when different components of the
discriminant locus intersect, or when a single component acquires a
singularity.  For a nonabelian gauge group factor $G_A$ that carries
no charged matter, it must be the case that the corresponding
divisor class $S_A$ has no intersection with the rest of the
discriminant locus $\Delta -\nu_A S_A$.  
(\ref{eq:6D-constraint-3}) and (\ref{eq:6D-constraint-4}) are simply
the conditions in the 6D supergravity theory that the divisor carrying
a gauge group $G_A$ in the F-theory picture have no intersection
either with the remainder of the discriminant locus or with any other
particular divisor carrying a gauge group.
Note that
\eq{eq:6D-constraint-4} is both necessary and sufficient for the
absence of charged matter under a given pair of gauge group factors
$G_A, G_B$, while \eq{eq:6D-constraint-3} is a necessary but not
sufficient condition for the absence of matter charged under a single
gauge group factor $G_A$.  Matter charged under a single gauge group
may appear for example as adjoint matter when $S_A$ is a higher-genus
surface, or from singularities within $S_A$ itself, as studied in
\cite{Sadov, Morrison-Taylor-matter}.
From the point of view of the low-energy theory, the relations
(\ref{eq:6D-constraint-3}) and (\ref{eq:6D-constraint-4}) are
apparently nontrivial constraints on the topological terms
\eq{GS-term} appearing in the action, although as mentioned above they
follow from anomaly cancellation conditions in six dimensions.  

Now we consider the constraint (\ref{eq:6D-constraint-5}), for which
we consider a 4D analogue in Section \ref{sec:constraints}.
In the F-theory picture any gauge group factor that carries no charged matter
is associated with a codimension one
singularity on a divisor with topology $\P^1$.  (Gauge groups on
divisors with higher genus topology always carry adjoint or equivalent
matter.)  We can use this fact to generalize the argument in the
previous subsection for the F-theory constraint associated with the
Euler character.  The correction to the Euler character of the total
resolved Calabi-Yau space $X$ when multiple gauge group factors $G_A$
arise on codimension one loci $S_A$ in the base surface $B$ gives
\cite{Klemm-lry}
\begin{equation}
 \chi (\hat X)  = -60 \int_B c_1^2 (B)  - \sum_A r_{G_A} c_{G_A} (2-2 g_A)\,,
\label{eq:withG_6D}
\end{equation}
where $r_{G_A}, c_{G_A}, g_{A}$ are the rank and dual Coxeter number of $G_A$, and
$g_A$ is the genus of $S_A$.  Since as noted above, $S_A$ is a genus 0 curve
when there is no matter charged under $G_A$, and $c_{G_A} r_{G_A} = {\rm dim}_{G_A}
-r_{G_A}$, the modification of the Euler character is just twice the difference
between the rank and the dimension of $G$.
The Euler character is then
\begin{equation}
\chi (X) = 60T-540+ 2 \sum_A (r_{G_A}- {\rm dim}_{G_A}) \label{eq:eval_withG_6D}
\end{equation}
where we have used \eqref{eq:6D-Euler} for the unmodified $\int c_1^2$
in \eqref{eq:withG_6D}.  Comparing \eqref{eq:eval_withG_6D} with the
general expression \eqref{generalchi_6D} one finds
\begin{equation}
29T -273 = \sum_A {\rm dim}_{G_A}-H \,,
\label{eq:constraint-5-again}
\end{equation}
thus producing from the F-theory geometry the constraint
(\ref{eq:6D-constraint-5}), equivalent to the anomaly constraint
\eq{eq:HVtext} for a theory with no charged matter fields.

Clearly, one can also perform the geometric transitions discussed in
\eqref{transition_map_6D} in an F-theory configuration with divisors
carrying nonabelian gauge groups. Using
\eqref{eq:withG_6D} and \eqref{changec1_6D}, the
change in the spectrum is identical to \eqref{Hodge_change}  as long
as the point blown up does not live in a divisor carrying a gauge
group factor.  
When the blown-up point lives on a divisor carrying a gauge group
factor, the gauge group generally changes, but this always occurs in a
fashion compatible with (\ref{eq:constraint-5-again})

\subsubsection{Sign constraints and the Kodaira condition}
\label{sec:signs-Kodaira}

In addition to the constraints on the spectrum that we have
already discussed,
F-theory imposes a set of positivity conditions on the vectors $a$ and
$b_A$
\begin{eqnarray}
j \cdot (-a) & > &  0 \label{eq:6D-constraint-ja}\\
j \cdot b &  > &  0\label{eq:6D-constraint-jb}\\
j \cdot (-12a-\sum_{A}\nu_A b_A ) & > & 0 \,. \label{eq:6D-constraint-jy}
\end{eqnarray}
The geometric statement of these conditions in F-theory is that the
anti-canonical class $-K$, all divisors $S_A$ carrying gauge group
factors, and the residual divisor 
locus $Y$ defined through
(\ref{eq:Kodaira}) are all effective divisors.

The condition (\ref{eq:6D-constraint-jb}) has a simple interpretation
in terms of the 6D supergravity theory; it states that the kinetic
term $F\wedge * F$ for each gauge group factor has the proper sign
\cite{Sagnotti}.  The other two conditions do not have known
interpretations in terms of the low-energy theory.
The condition (\ref{eq:6D-constraint-ja}) states that the quadratic
term in the curvature of the form $R \wedge * R$ must have a
specific sign.  As discussed in \cite{KMT-II}, this condition may
follow from causality, following an argument analogous to that of
\cite{Allan-Nima}.  We discuss the analogous 4D constraint in the
following section.

The Kodaira condition (\ref{eq:Kodaira})  from which
constraint (\ref{eq:6D-constraint-jy}) 
follows
is of
crucial importance in F-theory compactifications. 
The Kodaira condition expresses the geometric condition that the total
space $X$ is Calabi-Yau and hence preserves supersymmetry in the
dimensionally reduced theory.
In the weak coupling
limit this condition simply corresponds to the well-known fact that the 7-brane
tadpoles have to globally cancel.  
Written in cohomology
\eqref{eq:Kodaira} can be evaluated on a basis and amounts to
\begin{equation}
 -12 K^\alpha -\sum_{A} \nu_A C^\alpha_A - Y^\alpha = 0 \ .
\end{equation}
The geometric condition that the residual divisor locus $Y$ be
effective corresponds to the constraint (\ref{eq:6D-constraint-jy}).
It would be very interesting to achieve some understanding of this
constraint from the point of view of the supergravity theory.  In
particular, this inequality plays a key role in bounding the set of
possible F-theory compactifications when the number of tensors $T$
becomes large \cite{KMT-II}; understanding this constraint as a consistency
condition on low-energy theories would be one of the final steps
needed in matching low-energy consistency conditions to consistency
conditions from string theory for 6D theories \cite{universality}.

\subsubsection{Lattice structure for dyonic string charges}

As discussed earlier, in any supergravity theory arising from an
F-theory compactification, the lattice of dyonic string charges takes
the form $\Gamma = H_2 (B,\Z)$.  By Poincar\'{e} duality this lattice
is self-dual/unimodular.  Thus, F-theory imposes the constraint that
the dyonic string charge lattice is unimodular.  It was shown in
\cite{Seiberg-Taylor} that this condition is also necessary for any
consistent 6D supergravity theory.  Thus, this is an example of a
consistency condition arising from F-theory that is also a quantum
consistency constraint, where the understanding of the F-theory
picture motivated the identification of the macroscopic consistency
condition.  It may be that other F-theory constraints will eventually
be understood in this fashion from low-energy/macroscopic
considerations.

To understand the corresponding structure arising in four dimensions,
it may be helpful to review the nature of the inner product structure
on the lattice $\Gamma$.  In the F-theory picture this is just the
intersection form on $H_2 (B,\Z)$.  From the point of view of the
supergravity theory, the elements $x \in \Gamma$ are charges for
dyonic strings that couple to the self-dual and anti-self-dual
two-forms $B^\alpha$.  The inner product $x \cdot y$ of the charges
between two such dyonic strings  must be an integer by the
generalization of the Dirac quantization condition to six dimensions
\cite{Deser-quantum}.

\subsection{Geometrical constraints from F-theory on 4D  supergravity theories}
\label{sec:4D-constraints}

In this section we investigate various simple constraints that
the geometry of
F-theory imposes on 4D supergravity theories.  
While in 6D the constraints from F-theory are clear discrete
constraints on massless spectra that are satisfied across all continuous
branches of the moduli space, the constraints from F-theory on 4D
theories are more subtle.  In particular, as discussed above, in a
general 4D F-theory compactification many of the continuous geometric
moduli are lifted.  F-theory constraints on the spectrum become less
clear from the low-energy action when the fields become sufficiently
massive that the F-theory moduli are no longer clearly distinguishable
from other massive fields in the theory.  In the discussion here of
constraints on 4D supergravity theories, we assume that we are working
in a large volume compactification where fields coming from F-theory
geometry are all light and can be identified in the spectrum.
Understanding how these constraints and other structures extend
further into the moduli space beyond the large volume approximation
represents a challenge for future work.

Another significant limitation in treating F-theory constraints on 4D
theories is the absence of a systematic formalism for treating the
degrees of freedom on the 7-brane world volumes in a way that is
naturally compatible with Weierstrass models of F-theory.  In
particular, while recently there has been progress in understanding
certain classes of ``G-flux'' configurations in F-theory
\cite{Grimm-hkk}-\cite{Intriligator-jmmp}, and in principle fluxes can be integrated with F-theory
from the point of view of M-theory \cite{Denef:2008wq,Grimm:2011fx}, there is no
unified synthesis of gauge fluxes on 7-branes, or the related adjoint
scalars that appear in ``T-brane'' constructions \cite{T-branes} with
F-theory geometry.  This poses an obstacle to a systematic treatment
of constraints, particularly when matter fields are involved.  There
is a close interplay between fluxes, geometry and matter in 4D
theories; in particular, fluxes can change the matter content on a
7-brane world volume or at intersections between 7-branes.
We focus therefore on simple constraints from F-theory geometry that
do not depend critically on the detailed structure of matter.  A clear
direction for further extension of this work would be a more careful
treatment of matter, fluxes, and codimension two and three
singularities in 4D F-theory constructions.

\subsubsection{Constraints on theories without gauge groups}
\label{sec:constraints-no}

We begin our discussion as in 6D, by focusing on effective theories
that have no gauge group.  
Such theories arise, for example, for generic F-theory Weierstrass
models over Fano threefold bases such as $\P^3$ where the geometry
does not impose any gauge group on the theory by requiring vanishing
of $f, g$ on any particular divisor locus.
In this case, the associated
elliptically fibered Calabi-Yau fourfold has no singularities.
Moreover, the elliptic fibration has just a single section, so that
there are no massless $U(1)$ symmetries $r_{\rm v} = 0$, and we have
$\rp= 0$.  In such situations there is a constraint on the
spectrum analogous to \eq{eq:6D-scalar-constraint}.  The Euler character of the
elliptically fibered fourfold $X$ is related to the topology of the
base $\cB_3$ by \cite{Sethi:1996es,Klemm-lry}
\footnote{Note that the base $\cB_3$ of an elliptically fibered Calabi-Yau
  fourfold always has $24 \chi_0(\cB_3)=\int_{\cB_3} c_1(\cB_3) c_2(\cB_3) = 24$. This
  follows from the fact that $\chi_0(\cB_3)=\sum_n (-1)^n h^{0,n}(\cB_3) =1$,
  for a base of an elliptically fibered Calabi-Yau manifold since
  $h^{1,0}=h^{2,0}=h^{3,0}=0$ for $X$ and $\cB_3$.}
\begin{equation}
 \chi (X) = 288 + 360 \int c_1^3 (\cB_3) \,.
\label{eq:euler4d}
\end{equation}
Using \eqref{eq:general_euler} together with $r_{\rm v}=0$ this
gives the constraint
\begin{equation}
39 -60 \kappa_{\alpha \beta \gamma} K^\alpha K^\beta K^\gamma
= 39-60 \; \langle \langle a, a, a \rangle \rangle
= \csa + \cs-\cp\ ,
\label{eq:4D-scalar-constraint}
\end{equation}
on the numbers of the different types of scalar fields.  To streamline
equations and to clarify the analogy to 6D, we use here a shorthand
notation for the triple intersection product of three vectors under
$\kappa_{\alpha \beta \gamma}$, the intersection form on the base $\cB_3$
of a 4D F-theory compactification.
\begin{equation}
\langle \langle x, y, z \rangle \rangle \cong
\kappa_{\alpha \beta \gamma} x^\alpha y^\beta z^\gamma \,.
 \label{eq:triple-product}
\end{equation}
Note that this intersection product is computed using only the
components of $a^\alpha$ and not the component $a^0$ related to the axiodilaton.
The constraint (\ref{eq:4D-scalar-constraint}) should be
satisfied by any 4D ${\cal N} = 1$ supergravity theory arising from an
F-theory compactification in a phase with no unbroken gauge group.
A simple consequence of (\ref{eq:4D-scalar-constraint}) depends only
upon the light spectrum of the theory and not upon the details of
$\kappa_{\alpha \beta \gamma}$ or $K^\alpha$
\begin{equation}
 \csa + \cs-\cp \equiv 39 \; ({\rm
   mod}\ 60)\ .
\label{eq:4D-39-constraint}
\end{equation}

As a simple example of a 4D F-theory model satisfying
(\ref{eq:4D-scalar-constraint}), consider a generic Weierstrass model
over the base $\P^3$, with $\csa = h^{1, 1} (\cB_3) + 1 = 2$.
In this case, the cubic coupling of $a^\alpha$ is
characterized by a single integer 
\begin{equation}
\kappa =\kappa_{HHH}= 1 \,.
 \label{eq:k-p3}
\end{equation}
With $-K = 4 H$, in terms of the hyperplane section $H$,
the constraint (\ref{eq:4D-scalar-constraint}) then becomes
\begin{equation}
39 -60 \kappa a^3 = 3879
= 2 + \cs-(\cp + \rp) \,.
% \label{eq:}
\end{equation}
This is satisfied for a generic F-theory Weierstrass model on $\P^3$,
which has $\cs = h^{3, 1} (X) -1 = 3877$ scalar degrees of freedom,
and $\cp = \rp = 0$.  The quantity $h^{3, 1} (X)$ can be computed
directly for any toric base from the number of monomials in the global
Weierstrass model, minus the number of automorphisms.  The number of
automorphisms can be determined from the ``polar polytope'' \cite{Cox}
(for example see \cite{Denef:2008wq, mt-toric}).  

As another example of the constraint (\ref{eq:4D-scalar-constraint}),
consider the base $\tilde{\F}_2$.  Over this base there is no gauge
group required by vanishing of $f, g$ on any divisor.  From the form
of $-K = 2\Sigma + 5F$ and the triple intersection products given in
(\ref{eq:tfm-intersections}), we have
\begin{equation}
k = 2:\qquad  60\langle \langle -K, -K, -K \rangle \rangle = 3720,
% \label{eq:}
\end{equation}
and, using the fact that
for the base $\tf_2$, $h^{3,1} (X) = 3757$,
\begin{equation}
\csa + \cs-(\cp + \rp) = 3 + 3756-(0 + 0) =3759 =
39-60 \; \langle \langle a, a, a \rangle \rangle\,.
% \label{eq:}
\end{equation}
So (\ref{eq:4D-scalar-constraint}) is again satisfied.  

The base
$\tilde{\F}_3$ is an interesting case.  While $f, g$ are not required
to vanish on $\Sigma$ in this case, and there is no gauge group, there
is only a one-parameter family of constant functions for each of $f$
and $g$ that do not vanish on $\Sigma$.  In parallel with the 6D case,
where there is a $-2$ curve on $\F_2$ associated with a complex degree
of freedom that has been tuned and is not visible in the Weierstrass
parameterization \cite{mt-toric}, there is an extra degree of freedom
of type $\cp$ on $\tf_3$.  This combines with $\csa = 4357$ to give
\begin{equation}
\csa +\cs-\cp = 4359  \,,
% \label{eq:}
\end{equation}
again matching with \eq{eq:4D-scalar-constraint}.

The constraint \eq{eq:4D-scalar-constraint} is  reminiscent of
the analogous 6D constraint (\ref{eq:6D-constraint-1}) on the number
of scalars arising from the gravitational anomaly, although there is
no known pure gravitational anomaly in four dimensions.  
As we have emphasized repeatedly, in contrast to the 6D situation, 
constraints such as \eqref{eq:4D-scalar-constraint} and \eqref{eq:4D-39-constraint}
are only clearly formulated in the regime where the geometric moduli
of the F-theory compactification are light compared to other massive
fields in the theory, so that the numbers of scalar fields of each
type can be distinguished, and for (\ref{eq:4D-scalar-constraint}) so
that the canonical class $K$ and triple intersection form can be
extracted from couplings in the action as discussed in Section
\ref{sec:4D}.  To extend these constraints away from the class of
large-volume F-theory compactifications, it would be necessary to have
a definitive way of identifying $\cs,\csa,$ and $\cp$ from the point
of view of the low-energy theory in a general context, where some of
the fields may become very massive.  In the context of general ${\cal
  N} = 1$ supergravity theories, however, it is unclear how to make
sense of these moduli fields.  Not only do they mix with other massive
fields such as Kaluza-Klein modes, but they also can in principle mix
with one another, so that we do not have a definitive way of
distinguishing the fields $\cp$ from $\cs$ and $\csa$ away from the large-volume 
F-theory limit.  These considerations suggest that it may be difficult
to identify clear constraints that are valid for general 4D
supergravity theories not associated with a specific type of string
compactification.  Nonetheless, if any such global constraint on
${\cal N} = 1$ theories does exist,  constraints such as
(\ref{eq:4D-scalar-constraint}) that hold in
specific contexts such as F-theory should provide a helpful window and
guide to understanding the more general constraints.

Note that
in six dimensions, there are two constraints on the spectrum for
theories without gauge groups; in addition to
(\ref{eq:6D-constraint-1}), there is a second constraint
(\ref{eq:6D-constraint-2}).  It is natural to wonder whether there is
an analogous second constraint on the spectrum for 4D F-theory vacua without
unbroken gauge groups in four
dimensions.  We believe that there is no such second constraint, even
if we allow the number $\rp$ to enter the constraint, as suggested
by the structure of geometric transitions discussed below.  We
briefly outline the argument for this conclusion.  
Aside from the
spectrum, the only objects available for a constraint are
$\kappa_{\alpha \beta \gamma}$ and $K^\alpha$.  
The only invariant
that can be formed is the triple intersection $\langle \langle K, K, K
\rangle \rangle$.   If there is a second linear constraint then
there must be one linear combination that only contains the numbers
$C_*, \rp$ in the spectrum.  If there were such a linear
constraint on the spectrum then the existence of a pair of
compactifications that differ only in one number in the spectrum
would indicate that that number could not appear in the linear
combination.  As we discuss below, there exist such pairs,
indicating that $\cs$, $\csa$, $\rp$ cannot be in the linear
combination.  Since there are solutions with different values of
$\cp$ we conclude that there cannot be a further linear
constraint on the spectrum and triple intersection of $K$ for 4D
theories arising from F-theory compactifications.

\vspace*{0.05in}
\noindent {\bf Transitions between F-theory threefold bases}
\vspace*{0.05in}

Just as in the 6D story, transitions associated with blowing down
divisors in a threefold base 
connect different branches of the geometric moduli space of
elliptically fibered Calabi-Yau fourfolds that can be used to produce
a 4D supergravity theory from F-theory.  
Such transitions must respect the F-theory constraints on the
supergravity data such as the spectrum constraint described in the
previous section.  While in the physics of 4D
F-theory models some geometric
moduli are lifted by fluxes and the superpotential, these geometric
moduli still underlie the configuration space of the theory and
describe the off-shell geometry of the theory.  It is in this sense of
the underlying off-shell geometry that we can systematically describe
transitions between different F-theory geometries as connecting components of
the
continuous geometric moduli space, even though the physical moduli
space is more constrained.  Extending these off-shell parameters
outside the F-theory framework presents an interesting challenge for
developing a deeper understanding of the theory.

We can use invariance under these transitions as an aid in
understanding constraints on the physical spectrum of the theory.
The constraints may also shed light on the physics of the transitions.
For 4D theories, there is a much richer set of transitions than in 6D,
corresponding to different kinds of blow-up and blown-down processes.
The network of transitions for a particularly simple
class of (Fano) F-theory bases is explored in \cite{Grassi-network,
Mohri,
  Klemm-lry}.  The geometry of threefolds, however, is much more
complicated than that of surfaces.  
The mathematics of the Mori
program is aimed at understanding the connections between complex
varieties for dimensions 3 and higher analogous to minimal surface
theory in complex dimension 2, and classifying the types of
singularities that may arise \cite{Mori}.  A full exploration of the
physics associated with this
story will be a substantial research endeavor.  Here, we focus on the
simplest class of transitions, where a point or a curve in a smooth
base is blown up into a divisor in another smooth base, and where
neither base requires the presence of a gauge group.

Considering the blow-up of a single point in a smooth base
we can derive the change in the $\cN=1$ spectrum by
using the formula \eqref{eq:euler4d}. Note that in case
of a point blow-up the exceptional divisor is $\mathbb{P}^2$, which we
will refer to as $E$.
Using the formulae of appendix \ref{blowup_appendix}
together with \eqref{eq:euler4d},
the Euler characteristic obtained for an elliptic fibration over the
new base is
\begin{equation}
 \chi(X') = \chi(X) - 2880 \ .
\end{equation}
We can infer the change in the number of chiral multiplets by using
this equation. Note that $h^{1,1}(X)$ increases by one due to the new
exceptional divisor $E$. Since $E$ has no three-forms the Hodge number
$h^{2,1}(X)$ will not change. Hence, from (\ref{eq:general_euler})
the number of chiral multiplets
changes as
\begin{equation} \label{change_of_spec_1}
 \Delta \cs = - 481\ ,\qquad  \Delta \csa = +1 \ ,
\qquad  \Delta \cp = \Delta \rp= 0 \ .
\end{equation}
From the point
of view of the elliptic fibration the transition
\eqref{change_of_spec_1} can be viewed as a tuning of $481$ moduli to
enforce a singularity over the point in the base, which is then
resolved.  This number of moduli can also be derived directly by counting
degrees of freedom
in the Weierstrass model, as discussed below.
Note that the congruency condition \eq{eq:4D-39-constraint} is
invariant under this transition, since  $1-481 = -480 \equiv 0$ (mod
60).
This transition must be possible from the point of view of F-theory
geometry, but is not understood at this point physically from the
point of view of the low-energy theory.

As a simple example of a transition of this type, consider
$\tf_1$, which can be realized as
a blow-up of a point on $\P^3$ just as $\F_1$ is given by blowing up a
point on $\P^2$.  It is easy to check that under this blow up, the
numbers in the spectrum change through
(\ref{change_of_spec_1}).  
Indeed, in this case the change in the
number of complex structure moduli $h^{3, 1} (X)$ can be computed
directly along the lines of the argument in \cite{Witten-phase-mf}.
Starting from the F-theory base $\cB_3' =\tilde{\F}_1$, the divisor
$\Sigma$ can be blown down to give $\cB_3 =\P^3$.
We can describe  $\tilde{\F}_1$ as a $\P^1$ bundle over $\P^2$
in terms of
coordinates  $(x_1, x_2, x_3, x_4, x_5)$ subject to the relations
\begin{equation}
(x_1, x_2,
x_3, x_4, x_5)\sim
(\lambda x_1, \lambda\mu x_2,
\lambda x_3, \mu x_4, \lambda x_5) \,.
\label{eq:fmt-scaling}
\end{equation}
The functions $f, g$ are of degrees $(16, 8)$ and $(24, 12)$ in
$\lambda, \mu$.  Blowing down $\Sigma$ gives $\P^3$, where $\tilde{f},
\tilde{g}$ are of degree 16 and 24, and descend from $f, g$ as in the
case of $\F_1 \rightarrow \P^2$.  
We can then directly count the number of degrees of freedom that must
be tuned in $\tilde{f}, \tilde{g}$.  The power of $x_2$ in $f$ is at
most 8.  This requires tuning $1 + 3 + 6 + \cdots + 36 =120$
coefficients.  Similarly, $g$ has a power of $x_2$ that is at most 12,
requiring the tuning of $1 + 3 + \cdots + 78 = 364$ coefficients.
Thus 484 coefficients must be tuned, with a three-parameter space of
points where the tuning may be done, giving a total of 481 moduli that
are removed in the transition.  From the point of view of the theory
on the blown-down base $\cB_3$, the tuning just described corresponds
to arranging coefficients in the Weierstrass model so that there is a
codimension 3 point where $f, g$ vanish to degrees 8, 12.  This tuning
is local in the vicinity of the singularity that must be blown up.  It
is straightforward to verify by counting monomials in the dual
polytope (and subtracting automorphisms in the polar polytope as in
\cite{mt-toric}) that for any blow-up of a point in a toric base, the
number 481 of degrees of freedom that must be tuned to give a
transition of this type will be the same, as long as no additional
singularity (such as would give a gauge group or matter field) is
required on either threefold.  Note that a codimension 3 point where
$f, g$ vanish to degrees 4 and 6 is singular but cannot be blown up to
a divisor, suggesting a pathology of such theories that is not yet
well understood \cite{Morrison-codimension-3}.

In a three-dimensional base one can also blow up a smooth curve
$\cC$ to a divisor in the threefold base. 
This is the 4D analogue of the tensionless string transition in 6D
theories.  In the F-theory picture a 3-brane wrapped on a $\P^1$ fiber
over the curve $\cC$ in the blown up space becomes a tensionless
string in the limit as the fiber shrinks.
In
this case one finds a different pattern in the change of the Euler character
and spectrum.
The changes in the Chern classes and the intersection numbers of the blown-up
base are summarized in appendix \ref{blowup_appendix}. Using
\eqref{eq:euler4d} gives
\begin{equation}
  \chi(X') = \chi(X) - 1440\, \Big(2- 2 g_\cC + \int_\cC c_1(\cB_3) \Big)\ ,
\end{equation}
where $g_\cC$ is the genus of the curve $\cC$. In contrast to the
blow-up of a point, the blow-up divisor $E$ is a $\mathbb{P}^1$-bundle
over $\cC$
and hence has a more non-trivial topology. In particular, it has $g_{\cC}$ new
$(2,1)$-forms that are obtained by wedging the two-form of the
$\mathbb{P}^1$-fiber
with a $(1,0)$-form on $\cC$. This implies that  the number of
$(2,1)$-forms
of the base $\cB_3$ and the fourfold $X$ will also change in this transtion.
Translated into the change of the four-dimensional spectrum one
finds
\bea
 \label{change_of_spec_2}
 \Delta \cs &=& - 481 (1 -g_\cC) - 240 \int_{\cC} c_1(\cB_3) \
,\qquad  \Delta \csa = +1 \ , \\
 \Delta \rp &=& g_{\cC}\ , \qquad \Delta \cp = 0 \ . \nonumber
\eea
Note that this change in the spectrum will be modified if the blow-up
changes  the structure so that a divisor has $f, g$ vanishing,
requiring a change in the gauge group.  Even in the marginal case
where on some divisor $f, g$ are constant, as mentioned above for the example $\tf_3$,
additional scalars $\cs,\cp$ can arise in such a transition.

The change of spectrum in this transition includes a change in
$\rp$.  This
motivates us to generalize the constraint \eq{eq:4D-scalar-constraint}
to include models with $\rp$; including this term from
\eq{eq:general_euler} gives
\begin{equation}
39 -60 \kappa_{\alpha \beta \gamma} K^\alpha K^\beta K^\gamma
= 39-60 \; \langle \langle a, a, a \rangle \rangle
= \csa + \cs-\cp-\rp\ .
\label{eq:4D-scalar-constraint-r}
\end{equation}
Just as the change in spectrum under a tensionless string transition
is compatible with the constraints (\ref{eq:6D-constraint-1}),
(\ref{eq:6D-constraint-2}) we expect that the changes of spectrum
(\ref{change_of_spec_1}), (\ref{change_of_spec_2}) are compatible with
the 4D constraint (\ref{eq:4D-scalar-constraint-r}).  A detailed check
of this would require explicit computation of the triple intersection
numbers on both sides of the transition; this can be done in any
particular case from the geometry.  More generally, however, we can
easily confirm that the associated congruence 
\begin{equation}
 \csa + \cs-\cp-\rp \equiv 39 \; ({\rm
   mod}\ 60)\ .
\label{eq:4D-39-constraint-r}
\end{equation}
is invariant under both these transitions.  This serves as a check
that this constraint is indeed valid for general F-theory models.
Further transitions would be needed to span the space of
connected F-theory bases, however, and --- unlike in 6D --- in 4D not all
F-theory bases can be connected by the the transitions associated with
Mori theory: blowing up, blowing down, flips and flops.

We return now to complete the proof of the statement made in the
previous section that there is only one linear constraint involving
$\langle \langle K, K, K \rangle \rangle$ on the spectrum of fields of
4D F-theory models with $r_v= 0$.  Note that if we blow up a curve of
genus 0, the change in the spectrum is identical to that for blowing
up a point, except for the last term in $\Delta \cs$.  Since this term
can be nonzero, as discussed above there can be two compactifications
that only differ in this number in the spectrum, so $\cs$ cannot
appear in any linear constraint.  Given this, from
(\ref{change_of_spec_1}) it follows that $\csa$ also cannot appear.
But then $\rp$ also cannot appear since we can blow up a curve of
nonzero genus without changing $\rp$.  Since there are models with
different values of $\cp$ there cannot be any further linear
constraints beyond (\ref{eq:4D-scalar-constraint}).

\subsubsection{Constraints on 4D theories without charged matter}
\label{sec:constraints-group}

We now relax the condition that there is no gauge group in the 4D
theory, and generalize the 4D constraint derived above to include
theories with gauge groups.
Including charged matter in
4D is significantly more involved compared to 6D.  
Unlike in 6D, where codimension two loci in the base giving matter are
pointlike, in 4D codimension two singularities are themselves
surfaces, whose Euler character affects the matter content of the theory.
Fluxes $G_4$ alter
the spectrum by modifying the equations for the massless matter
eigenstates on the worldwolume and intersections of the 7-branes.  
In this paper we only comment briefly on the
complications associated with chiral matter and fluxes, and focus on
topological constraints on $\cB_3$ and $X$.
There are presumably more complicated constraints involving charged
matter fields when fluxes are correctly included, perhaps related to
the relation (\ref{eq:22-constraint}).

As in six dimensions,
a simple condition that implies the absence
of any matter charged under two groups $G_A$ and $G_B$ associated with
divisors $S_A, S_B$ is that the intersection between the two divisors
vanish identically
\begin{equation}
\kappa_{\alpha \beta \gamma} S_A^\alpha S_B^\beta = 0, \; \forall
\gamma \,.
\label{eq:4D-ss-constraint}
\end{equation}
This is parallel to the 6D constraint \eq{eq:6D-constraint-4}, though
in the 4D case this is only a sufficient condition for the absence of
multiply charged matter while in 6D the condition is also necessary,
since in 6D every intersection between divisors is a pointlike
codimension two singularity that carries matter degrees of freedom.
On the other hand, the 6D condition (\ref{eq:6D-constraint-3}) stating
that the divisor carrying a gauge group is orthogonal to the residual
divisor locus $Y$
is
necessary for theories without matter, but not sufficient since a
gauge group can carry non-local matter such as an adjoint.  In 4D the
analogous condition is neither necessary or sufficient, for the same
reasons stated above.

The constraints on the absence of 4D chiral matter have a closer 
analogy to the 6D constraints of section \ref{sec:6D-without}. Physically this 
is due to the fact that 4D chiral matter induces non-Abelian anomalies, 
just as a general matter spectrum does in 6D. We have recalled 
in \eqref{res_Theta} that the chiral spectrum of a 4D F-theory compactification 
on the singular space $X$ can be derived from the $G_4$ flux 
on $\hat X$ by studying the constant couplings $\Theta_{i_A i_B}$. 
These couplings capture the 4D chiral spectrum as 3D loop corrections in 
the Coulomb branch.   
Thus, absence of 4D chiral matter simply implies 
\beq \label{constraint_chiral}
   \Theta_{i_A i_B} = \int_{\hat X} \omega_{i_A} \wedge \omega_{i_B} \wedge G_4 = 0\ ,
\eeq
where $\omega_{i_A}$, $\omega_{i_B}$ are the resolution divisors for the gauge 
groups $G_A,G_B$ in the M-theory picture. We note that it is possible to have $A=B$ in \eqref{constraint_chiral}, 
which captures information about the chiral matter of the intersection of $S_A$ with the 
rest of the discriminant. This implies that each
F-theory compactification without chiral matter has to admit a special non-trivial $G_4$ satisfying 
\eqref{constraint_chiral} together with 
the D3-tadpole cancellation condition \eqref{D3-tadpole}. The 
6D analog of \eqref{constraint_chiral} is a constraint on the triple intersection numbers 
of the resolved Calabi-Yau threefold with three indices labeling exceptional resolution 
divisors for the gauge groups. In order to promote \eqref{constraint_chiral} to a constraint
on the low-energy data, just as in 6D one has to find other couplings that involve the 
same flux data. It will be interesting to find such couplings in 4D compactifications. 
The immediate analog to 6D appears if a 4D Green-Schwarz coupling is required 
to cancel anomalies. 

For theories in which there is no charged matter and there are no
codimension two singularities on the gauge group divisor loci, the
argument leading to (\ref{eq:4D-scalar-constraint}) can be generalized
to include theories with gauge group factors.  A set of identities
known as Pl\"ucker identities can be used to show that the Euler
character of the Calabi-Yau fourfold that is elliptically fibered over
a threefold base with a homogeneous degeneration over divisors
carrying a pure gauge group with no codimension two singularities
is given by \cite{Klemm-lry, Andreas-Curio}
\begin{equation}
 \chi(X) = 288 + 360 \int_B c_1(B)^3 - \sum_A  r_{G_A} c_{G_A}
(c_{G_A}+1) \int_{S_A} c_1(S_A)^2 \,.
\end{equation}
This generalizes the constraint (\ref{eq:4D-scalar-constraint}) for a
theory with pure gauge group factors and no matter to
\begin{eqnarray}
\lefteqn{39-60 \; \langle \langle a, a, a \rangle \rangle}\nonumber \\
 & = &  \csa + \cs + r_v-(\cp + \rp) + \frac{1}{6}  \sum_A  r_{G_A} c_{G_A}
(c_{G_A}+1) \langle \langle a + b_A,a + b_A,b_A \rangle \rangle\,.
\label{eq:4D-constraint-g}
\end{eqnarray}
As an example, consider $\tf_4$, which carries a gauge group $SU(2)$
on the divisor class $\Sigma$, as discussed above.  In this case we have
$-K = 2 \Sigma + 7F$, $h^{3, 1} =  5187$, and the group has rank
$r_v= 1$  so
\begin{equation}
39-60 \langle \langle a, a, a \rangle \rangle
=5199= 3 + 5186 + 1+\frac{1}{6} (1\cdot 2 \cdot  3 \cdot 9) \,,
% \label{eq:}
\end{equation}
confirming \eq{eq:4D-constraint-g}.

\subsubsection{Sign conditions and Kodaira condition}

Just as in six dimensions, the anticanonical class $-K$ and the
divisors $S_A$ carrying any nonabelian gauge group must be effective.
This imposes, in particular, positivity constraints on the volumes of
these divisors, so that in the 4D theory we must have
\begin{equation}
\langle \langle -K, v, v \rangle \rangle =
\kappa_{\alpha \beta \gamma} (-a^\alpha) v^\beta v^\gamma > 0
\label{eq:4D-a-positive}
\end{equation}
and
\begin{equation}
\langle \langle S_A, v, v \rangle \rangle =
\kappa_{\alpha \beta \gamma} b_A^\alpha v^\beta v^\gamma > 0 \,.
\label{eq:4D-b-positive}
\end{equation}
The analogous constraint in 6D to (\ref{eq:4D-b-positive}) simply
corresponds in the supergravity theory to the constraint that the
kinetic term for the gauge group factor $G_A$ has the proper sign.  In
4D this is complicated by the appearance of the additional real second
term in (\ref{Talpha-definition}) and additional axions discussed in section 
\ref{additional_axion}, which as noted below (\ref{eq:triple-product}) are not
included in the intersection product $\langle \langle \cdot, \cdot,
\cdot \rangle \rangle$.

As discussed in the 6D context, the Kodaira constraint
\eq{eq:Kodaira} is another constraint imposed by F-theory that at the
present time is not clearly understood from the low-energy point of
view.  
The Kodaira constraint sharpens the inequalities \eq{eq:4D-a-positive}
and \eq{eq:4D-b-positive} to
\begin{equation}
12\langle \langle -K, v, v \rangle \rangle
\geq \sum_{A}\nu_A \langle \langle S_A, v, v \rangle \rangle > 0 \,.
\label{eq:4D-Kodaira}
\end{equation}

As an example of the Kodaira constraint, consider again F-theory
compactifications on the base $\P^3$, with a gauge group $SU(N)$.
In this case, as in (\ref{eq:k-p3}), there is only a single axion
$\rho_x$ other than $\rho_0$, and $-a = 4H$.  The $SU(N)$ gauge group
lives on a divisor that can be identified from the 4D
Green-Schwarz-like couplings to be $b = mH$, with $m > 0$ an integer.
The Kodaira constraint in this case is
\begin{equation}
48 \geq m N \,.
% \label{eq:}
\end{equation}
In \cite{Morrison-Taylor-matter} a more careful analysis of such theories
with $m = 1$ ($b = H$) showed that in this case $N \leq 32$.

We do not have any clear understanding at present of how the
inequalities (\ref{eq:4D-a-positive}) and (\ref{eq:4D-Kodaira})
should be understood in terms of the low-energy theory.
We note in passing, however, that some possibly related constraints on
Gauss-Bonnet couplings have been discussed using AdS/CFT e.g.~in
\cite{Buchel}.

\subsubsection{Lattice structure for string states}
\label{sec:4D-lattice}

In 6D supergravity theories, the charges $a, b_i$ appearing in the
$B R^2, BF^2$
topological couplings live in a sublattice $\Lambda$
of the lattice of dyonic charges $\Gamma$ associated with
strings in space-time.  As discussed in the 6D section,
there is an inner product on $\Gamma$
associated with Dirac quantization, under which  $\Gamma$ takes the
structure of an integral and self-dual (unimodular) lattice.  This
lattice plays an important role in the relationship with F-theory; the
charges on this lattice characterize elements of $\htb$.

There is a similar lattice structure in 4D, though the absence of an
inner product makes the story less transparent.
In 4D, the classical continuous axionic shift symmetries are broken to
a discrete lattice $L$, so that we have an invariance under
\begin{equation}
\rho_\alpha \rightarrow \rho_\alpha + l_\alpha, \; \; l \in L  \,.
\label{eq:axion-shift}
\end{equation}
This breaking of the continuous shift symmetries arises from
nonperturbative terms.
This can also be
understood in terms of quantized strings that carry magnetic axion
charges $q$ in the  lattice $L$.  A quantum string gives rise to an axionic
charge measured along a loop around the string where $\int d
\rho_\alpha = l_\alpha$ is an element of $L$.
The charges $a, b_A$ that
parameterize the 4D topological couplings (\ref{eq:4D-topological})
must lie in the dual lattice
\begin{equation}
a, b_A \in L^*\,,
% \label{eq:}
\end{equation}
since a shift of the axions under (\ref{eq:axion-shift}) must leave
the action invariant up to an overall additive constant $2 \pi  k, k
\in\Z$, and the gauge and gravitational instanton numbers are
integrally quantized.
From the F-theory point of view, this characterization follows from
the fact that axions are associated with components of $C_4$ that are
4-forms in $\cB_3$, while $a, b_A$ are associated with divisor classes
in $H_4 (\cB,\Z)$.

For a 4D F-theory compactification, the lattice $L$ characterizes
the charges of fundamental strings in space-time that are charged
under the axion $\rho_0$ as well as axionic strings that arise from
D3-branes wrapped on cycles in $\htb$.  These strings are
electrically charged under the two-form fields $B^\cA$ and
magnetically charged under the axions $\rho_\cA$.  Whereas in 6D,
the intersection product between dyonic strings has a clear physical
interpretation in terms of the phase appearing in the Dirac
quantization condition, we do not have an analogous simple physical
interpretation of any structure associated with a set of 3 strings or
instantons in four dimensions.  The structure of the triple
intersection product suggests that there is a natural triple product
between sets of 3 axionically charged strings in the 4D theory.  This
may be more naturally formulated in the language of instantons.
Euclidean D3-branes wrapped on 4-cycles $H_4 (B,\Z)$ correspond to
instantons in the 4D theory that couple to the axions $\rho_\alpha,
\alpha > 0$.  The triple intersection product $\kappa_{\alpha \beta
  \gamma}$ naturally associates an integer with any set of 3 such
instantons.  It would be interesting to identify a natural physical
interpretation of this product in four dimensions.

In principle, the set of supersymmetric axionically-charged string
excitations of a 4D theory describes the {\it Mori cone} of effective
curves on the F-theory compactification threefold $\cB_3$.
This characterizes the complex structure of the compactification space.
The geometric F-theory structure of axions and axionically charged
strings/instantons in 4D is complicated, however, by extra 4D axions
as discussed above.
In parallel to the additional axion from the axiodilaton, 
additional instantons appear in four dimensions 
associated with pointlike D(-1)-branes; these
instantons couple to $\rho_0$ just as Euclidean D3-branes couple to
the other axions $\rho_\alpha, \alpha > 0$.  As discussed in Section
\ref{sec:4D-heterotic}, the 4D axions have a different geometric
interpretation in heterotic compactifications, and the axion $\rho_0$
ties into the geometric structure of the compactification manifold.
Away from the large-volume F-theory limit, many of the axion fields
become massive and their identity becomes difficult to distinguish.
It seems likely that in general 4D supergravity theories, the
axion-instanton couplings can be completely general; gauge and
gravitational instantons may couple to any axions in the theory,
including those in $\cp$ and $\cs$ as well.  In this case
there will be massive string excitations associated with each of the
axions.  Many of these theories will not have large-volume F-theory or
heterotic interpretations.  A better understanding of additional
structure such as the triple intersection product on the axionic
string/instanton charge lattice may be of value in developing this
part of the story further.

\section{Conclusions}
\label{sec:conclusions}

We have found that, just as in six dimensions, much of the geometric
data associated with an F-theory compactification to four dimensions
is encoded in the spectrum and action of the 4D theory.  In
particular, for large volume compactifications the spectrum of light
fields directly encodes the Hodge numbers of the F-theory
compactification manifold, and the canonical class and 7-brane divisor
classes on the F-theory base are encoded in topological terms coupling
4D axions to curvature squared terms in the action.  
Strings carrying magnetic axion charges  and couplings in the
supergravity action further characterize the full second homology and
triple intersection product of the F-theory compactification manifold.
In 6D this correspondence makes it possible to read off the F-theory geometry
directly from the structure of the low-energy theory.  In 4D, this
correspondence is only transparent in the large-volume limit where the
geometric moduli are light.  F-theory geometry in 4D is  obscured
in the low-energy theory in the bulk of the moduli space due to
various corrections, lifting of moduli by the superpotential generated
by fluxes,  additional axions, and various other
complications.  Further analysis of how the results of this work can
be relevant outside the large-volume F-theory
limit is an interesting direction for
future work.

A particular element that has played a key role in understanding the
space of 6D supergravities, and that seems to play a related
structural role in 4D supergravities, is the set of couplings between
two-form/axion fields and curvature squared terms in the action.  In
4D, axions have a discrete shift symmetry associated with magnetic
charges of stringlike excitations of the theory.  These axions couple
to $F \wedge F$ and $R \wedge R$ in the 4D action in a way that
captures topological aspects of the compactification geometry in the
case of models that arise from an F-theory compactification.  These
coupling terms illuminate the structure of 4D supergravity theories
just as Green-Schwarz terms illuminate 6D supergravity theories, even
though in 4D  these terms are not strongly restricted by anomalies.  One
specific application of these terms that we have explored in this
paper is to heterotic/F-theory duality.  We have computed the
axion--curvature squared terms explicitly in a general class of 4D
heterotic compactifications with F-theory duals, providing a check on
the general structure presented here, 
and giving a simple dictionary
indicating which $\P^1$ bundle acts as the F-theory base and
which divisors in an F-theory compactification carry the
gauge group factors in any case of heterotic/F-theory duality.  These
results provide a complementary perspective to other methods such as
the spectral cover method and the stable degeneration limit 
\cite{fmw, Bershadsky-jps, Curio-Donagi}
previously used for understanding heterotic/F-theory duality in four
dimensions, and suggest that further insight into this duality may
follow from further considerations along the lines pursued in this
paper.  For example, the axion--curvature squared terms provide
information about heterotic/F-theory duality for the $SO(32)$ theory
that is less amenable to analysis through the stable degeneration limit.
Further development of heterotic/F-theory
and other dualities through the structure of the low-energy theory and
topological axion-curvature squared couplings may
help to clarify how the various string constructions of 4D
supergravity theories are related and to chart the space of 4D ${\cal
  N} = 1$ string compactifications.

We have identified some constraints on 4D supergravity theories that
hold in the large-volume F-theory limit where the geometric moduli of
the compactification remain light.  The simplest
of these constraints is a condition on a linear combination of the
numbers of different types of fields (\ref{spectrum-split}) in the theory 
\begin{equation}
 \csa + \cs -\cp  \equiv 39 \; ({\rm
   mod}\ 60)\ ,
\label{eq:conclusions-39}
\end{equation}
for any theory with a completely broken gauge group.
This
constraint is an analogue of the gravitational anomaly constraints
appearing in 6D supergravity theories, though there is no known
gravitational anomaly in 4D that would give rise to constraints of
this form.  A more precise version of this constraint involves the
canonical class $K$ of the F-theory base, which is encoded in the
axion-curvature squared terms .
Note that a consequence of (\ref{eq:conclusions-39}) is that any
large-volume F-theory model must have at least 39 light scalar fields,
independent of the distribution of fields between the various types.
Both in 6D and in 4D there are also constraints from F-theory on the
signs of curvature-squared terms proportional to $R \wedge{}^*R$ and
$F \wedge{}^*F$.  In the latter case this constraint simply follows in
the low-energy theory from the condition that the gauge kinetic term
have the standard (negative) sign for stability of the theory.  In the
case of the metric curvature terms, no low-energy reason for a sign
constraint is known.

A key issue that should be incorporated better into the considerations
of this paper regarding 4D F-theory vacua is the role of fluxes, and
more generally world-volume fields on the 7-branes.  While F-theory
provides a simple and beautiful context for nonperturbative
exploration of a large region of the space of possible string
compactifications, this formulation of string theory is still
incomplete.  In particular, the description in terms of a Weierstrass
model is not coupled to a natural description of gauge fields on the
world volume, or fields such as the adjoint scalars in the Higgs
bundle on 7-branes.  Fluxes, however, play a key role in determining the
physics of ${\cal N} = 1$ 4D string vacua.  While these fluxes are
conceptually clear in the M-theory picture, this
framework loses the geometric picture of the Weierstrass model in the
F-theory context.  As a result, the tools for working simultaneously
with F-theory Weierstrass models and fluxes are still at an early
stage of development.  To understand the kinds of constraints we have
discussed here better, the incorporation of fluxes is clearly crucial.
In particular, fluxes play a key role in determining the structure of
matter in 4D theories.  We leave the further integration of fluxes into the
story begun here as a challenging open problem for future research.

In six dimensions, the lattice of dyonic string charges plays an
important role in the structure of an ${\cal N} = 1$ supergravity
theory.  
The structure of axion-curvature squared terms in 4D supergravity
theories arising from F-theory suggests that the analogous lattice of
axionic string charges should play a similar role in four-dimensional
theories.  
Away from the large-volume F-theory limit,
it seems that all axions in the theory can mix, but some integral
structure will still be supplied by the underlying massive axionic
string lattice. In particular, for any massive string state the associated magnetic
charge will correspond to an axionic shift symmetry.  Even when supersymmetric
string states are very massive, they still play a role in the basic symmetry
structure of the theory.
A better physical understanding of the 
axionic string lattice in low-energy ${\cal N} = 1$ supergravity
theories may be a key to applying the methods and results of this
paper to deeper issues in the structure of 4D theories.
It may be that further consideration of the world-volume theory on the
charged strings will shed light on constraints and/or structure in
general 4D supergravity theories.

A central question in the study of string compactifications
is the extent to which the geometry of the compactification can be
uniquely identified from data in the supergravity theory.  In six
dimensions, the story in this regard is quite clear.  In many cases,
knowledge of the spectrum and Green-Schwarz coefficients of the
low-energy supergravity theory is sufficient to uniquely determine the
geometry of a corresponding F-theory construction, when one exists.
In all cases, further knowledge of the spectrum and Dirac-quantized
charge products of supersymmetric dyonic string states fixes the
intersection product and Mori cone of the base, uniquely determining
the F-theory geometry when it exists.  In four dimensions, the story
is more complicated.  Only in the large-volume regime of F-theory are
the geometric moduli of the F-theory compactification clearly
distinguishable from other massive modes in the theory.  In this
regime the couplings between axions and curvature-squared terms play a
similar role to the Green-Schwarz terms in six dimensions.  Combining
this information with the string charge lattice, triple intersection
product, and other information from the supergravity spectrum and
action it is in principle possible to determine a corresponding
F-theory geometry. Beyond the lifting of moduli by fluxes and other
effects, however, there are also additional fields, such as the axion
that is associated with the axiodilaton in the weak coupling limit,
which must be disentangled in order to identify the F-theory geometry.
While in the case of heterotic/F-theory duality we were able to
explicitly identify the correspondence between axions to determine the
map between heterotic and F-theory geometry, it is not clear how this
can be done in general.  It can be, for example, that there are dual
F-theory compactifications on distinct threefold bases that give
equivalent 4D physics, but with a different distribution of axions
between complex structure moduli and other fields.  For example, this
may occur if a Calabi-Yau fourfold admits two distinct elliptic
fibrations with different bases. Situations of this
kind in the dual heterotic setting have been discussed, for example,
in \cite{Distler-Kachru, Blumenhagen-r}.  Further analysis of the
extent to which 4D supergravity determines compactification geometry
promises to lead to new insights into the global structure of the
space of string vacua, for which the tools and methods developed in
this paper may be of some use.

Another direction in which the methods of this paper may be
applied is towards the systematic understanding of the space of
elliptically fibered fourfolds underlying the space of 4D F-theory
models. Just as for elliptically fibered threefolds, elliptically
fibered fourfolds form a complicated moduli space, with continuous
branches connected by non-Higgs type phase transitions such as
tensionless string transitions.  Recently, a global exploration of the
space of threefold bases for 6D F-theory vacua has been initiated
\cite{Morrison-Taylor, mt-toric}, following the minimal model approach
\cite{Grassi}.  The structure of the axion-curvature squared terms and
constraints described in this work may provide useful tools in
exploring the space of 4D F-theory vacua, and a better understanding
of the connections between the branches of the theory associated with
different bases and the exotic transitions connecting these theories.

\noindent
{\bf Acknowledgements}: We would like to thank Allan Adams, Lara
Anderson, Ralph Blumenhagen, Federico Bonetti, Michael Dine, Michael
Douglas, Dan Freedman,
Jonathan Heckman, Stefan Hohenegger, Shamit Kachru, Denis Klevers,
Vijay Kumar, Hong Liu, Joe Marsano, John McGreevy, James McKernan,
David Morrison, Daniel Park,
Raffaele Savelli, and Timo Weigand  for discussions.
Thanks to the Simons Center for Geometry and Physics for hospitality
during the initiation and completion of this work.  This research was
supported in part by the DOE under contract \#DE-FC02-94ER40818.

\appendix

\section{Blowing up curves and points in a smooth threefold base}
\label{blowup_appendix}

In this appendix we summarize the necessary equations to discuss the blow-ups of points and smooth curves 
in a threefold $B$. We denote the blown-up space by $B'$ and name the blow-down map $\pi: B' \rightarrow B$. 
The exceptional divisor obtained after blow-up is denoted by $E$.

We first consider the blow-up of a point in $B$. The exceptional divisor $E$ in this 
blow-up is simply a $\mathbb{P}^2$.
The Chern classes of the threefold change as \cite{Griffiths}
\begin{equation}
   c_1(B') =  \pi^* c_1(B) - 2[E]\ ,\qquad c_2(B') = \pi^* (c_2(B)) \ , 
\end{equation}
We will also need the intersection numbers after blow-up. One finds that 
\begin{equation}
  E^2 = f \ , \qquad E\cdot f = -1 \ , \qquad E \cdot \pi^* D =  E \cdot \pi^* \tilde C = f \cdot \pi^* D =0 \ .
\end{equation}
for all divisors $D$ and curves $\tilde C$ in $B$. 

Let us now turn to the blow-up of a smooth curve $\cC$. The exceptional 
divisor is then given by the projectivisation of the normal bundle $N_B\cC$ of the 
curve in $B$, i.e.~$E = \mathbb{P}(N_B \cC)$.
The Chern classes of $B$ now change as 
\bea
   c_1(B') &=&  \pi^* c_1(B) - [E]\ ,\\
   c_2(B') &=& \pi^* (c_2(B) + [\cC]) - [E] \wedge \pi^* c_1(B)  \ , 
\eea
The blow-up space has the following intersection numbers 
\bea
  E^2 &=& - \pi^* \cC + \text{deg}(N_B \cC) f\ , \qquad E\cdot f = - 1\ ,\\
  E \cdot \pi^* D &=& (C\cdot D) f \ , \qquad f \cdot \pi^* D = 0 \ ,\qquad E \cdot \pi^* \tilde \cC = 0 \ ,
\eea
for all divisors $D$ and curves $\tilde C$ in $B$. The degree of $N_B\cC$ can be written as
\begin{equation}
  \text{deg}(N_B \cC) = -\chi(\cC) + \int_\cC c_1(B)\ .
\end{equation}

\section{Anomalies in 6D supergravity} \label{6Danomalies}

The anomaly
cancellation condition can be written in terms of the 8-form anomaly
polynomial as \cite{gsw, Sagnotti, Erler, Sadov}
\begin{equation}
I_8(R,F) = \half \Omega_{\alpha\beta} X^\alpha_4 X^\beta_4.
\label{eq:factorized-anomaly}
\end{equation}
Here
\begin{equation}
X^\alpha_4 = \half a^\alpha \tr R^2 +  \sum_i b_i^\alpha \
\left(\frac{2}{\lambda_i} \tr F_i^2 \right)
\label{eq:string-current}
\end{equation}
with $a^\alpha, \ b_i^\alpha$ transforming as vectors in the space
$\R^{1,T}$ with symmetric inner product $\Omega_{\alpha\beta}$;
``$\tr$'' of $F_i^2$ denotes the trace in the fundamental representation, and
$\lambda_i$ are normalization constants depending on the type of each
simple group factor.  Cancellation of the individual terms in
\eq{eq:factorized-anomaly} gives
\begin{eqnarray}
H-V & = &  273-29T\label{eq:hv}\\
0 & = &     B^i_{\rm adj} - \sum_{\bf R}
x^i_{\bf R} B^i_{\bf R} \label{eq:f4-condition}\\
a \cdot a & =   &9 - T  \label{eq:aa-condition}\\
-a \cdot b_i & =  & \frac{1}{6} \lambda_i  \left(  \sum_{\bf R}
x^i_{\bf R} A^i_{\bf R}-
A^i_{\rm adj} \right)  \label{eq:ab-condition}\\
b_i\cdot b_i & =  &\frac{1}{3} \lambda_i^2 \left(  \sum_{\bf R} x_{\bf
  R}^i C^i_{\bf R}  -C^i_{\rm adj}\right)  \label{eq:bb-condition}\\
b_i \cdot b_j & = &  \lambda_i \lambda_j \sum_{\bf R S} x_{\bf R S}^{ij} A_{\bf R}^i
A_{\bf S}^j\label{eq:bij-condition}
\end{eqnarray}
where  $A_{\bf R},
B_{\bf R}, C_{\bf R}$ are group theory coefficients defined through
\beq
\tr_{\bf R} F^2  = A_{\bf R}  \tr F^2 \ , \qquad \quad 
\tr_{\bf R} F^4  = B_{\bf R} \tr F^4+C_{\bf R} (\tr F^2)^2 \label{eq:bc-definition}\,,
\eeq
and where
$x_{\bf R}^i$ and $x_{\bf R S}^{ij}$
denote the number of matter fields that transform in the irreducible
representation ${\bf R}$ of gauge group factor $G_i$,
and $({\bf R} , {\bf S})$ of $G_i \otimes G_j$ respectively.
Note that for groups such as $SU(2)$ and $SU(3)$, which lack a fourth
order invariant, $B_{\bf R} = 0$ and there is no condition
\eq{eq:f4-condition}.

It is shown in \cite{KMT-II} using elementary group theory that the
inner products on the LHS of conditions
(\ref{eq:aa-condition}-\ref{eq:bij-condition}) are all integral as a
consequence of global and local anomaly cancellation.  This gives an
integral {\em anomaly lattice} $\Lambda$ formed from vectors $a, b_i
\in\R^{1, T}$.  The vector $a$ is associated with a coupling $a \cdot
B \; \tr R^2$ of the $B$ fields to space-time curvature, while the
vectors $b_i$ are associated with couplings $b_i \cdot B\;\tr F_i^2$
of the $B$ fields to the field strengths $F_i$ of the various factors
in the gauge group; together these terms form the Green-Schwarz
counterterm
\begin{equation}
B \cdot X_4 = B^\alpha
 \Omega_{\alpha\beta}
\left[\half a^\beta \tr R^2 +  \sum_i b_i^\beta \
\left(\frac{2}{\lambda_i} \tr F_i^2 \right)\right]
\label{eq:Green-Schwarz-counter}
\end{equation}
The lattice $\Lambda$ is a sublattice of the
complete unimodular lattice $\Gamma$ of dyonic string charges for the
6D theory.

\end{document}